\documentclass[aps,prd,twocolumn,amsmath,amssymb,amsfonts,nofootinbib,superscriptaddress]{revtex4-1}
\usepackage[utf8]{inputenc}
\usepackage{graphicx}% Include figure files
\usepackage{subfigure}
\usepackage{datatool}
\usepackage[scientific-notation=true]{siunitx}
\usepackage[dvipsnames]{xcolor}
\usepackage{lineno}
\usepackage{xspace}
\usepackage{orcidlink}
\usepackage{url}

\usepackage{breakurl}
\usepackage{lineno}
\def\gwh{gravitational-wave\xspace}
\def\cwh{continuous-wave\xspace}
\def\gws{gravitational waves\xspace}

\def\et{Einstein Telescope\xspace}
\def\ce{Cosmic Explorer\xspace}
\def\dm{dark matter\xspace}
\def\dmh{dark-matter\xspace}
\def\pbh{primordial black hole\xspace}
\def\pbhs{primordial black holes\xspace}

\def\bbhh{binary black-hole\xspace}

\def\bnsh{binary neutron-star\xspace}

\def\snr{signal-to-noise ratio\xspace}

\def\GFH{Generalized frequency-Hough\xspace}

\def\fivehzimprove{50\xspace}
\def\eighthzimprove{25\xspace}

\newcommand{\bea}{\begin{eqnarray}}
\newcommand{\eea}{\end{eqnarray}}
\newcommand{\be}{\begin{equation}}
\newcommand{\ee}{\end{equation}}

\newcommand{\avgVT}{\left\langle VT \right\rangle}

\newcommand{\dd}{\ensuremath{\mathrm{d}}}
\newcommand{\diff}[2]{\ensuremath{\dfrac{\dd {#1}}{\dd {#2}}}}

\newcommand{\TFFT}{T_\text{FFT}}

\newcommand{\Tobs}{T_\text{obs}}
\newcommand{\fpbh}{f_\text{PBH}}
\newcommand{\ftilde}{\tilde{f}}
\newcommand{\fsup}{f_\text{sup}}
\newcommand{\Mc}{\mathcal{M}}
\newcommand{\tew}{T_\text{EW}}
\newcommand{\tcoal}{t_\text{coal}}
\newcommand{\fmin}{f_\text{min}}
\newcommand{\fmax}{f_\text{max}}

\def\erfc{\mathrm{erfc}}

\newcommand{\tcb}{\textcolor{blue}}

\begin{document}

\title{Enabling multi-messenger astronomy with continuous gravitational waves: \\ early warning and sky localization of binary neutron stars in Einstein Telescope }

\author{Andrew L. Miller\,\orcidlink{0000-0002-4890-7627}}
\email{andrew.miller@nikhef.nl}
\affiliation{Nikhef -- National Institute for Subatomic Physics,
Science Park 105, 1098 XG Amsterdam, The Netherlands}
\affiliation{Institute for Gravitational and Subatomic Physics (GRASP),
Utrecht University, Princetonplein 1, 3584 CC Utrecht, The Netherlands}
\author{Neha Singh}
\email{neha.singh@uib.es}
\affiliation{Departament de F´ısica, Universitat de les Illes Balears, IAC3–IEEC, E-07122 Palma, Spain}
\affiliation{Laboratoire d’Annecy-le-Vieux de Physique Théorique (LAPTh), USMB, CNRS, F-74940 Annecy, France}
\author{Cristiano Palomba}
\email{cristiano.palomba@roma1.infn.it}
\affiliation{INFN, Sezione di Roma, I-00185 Roma, Italy}

\date{\today}

\begin{abstract}

Next-generation gravitational-wave detectors will provide unprecedented sensitivity to inspiraling binary neutron stars and black holes, enabling detections at the peak of star formation and beyond. However, the signals from these systems will last much longer than those in current detectors, and overlap in both time and frequency, leading to increased computational cost to search for them with standard matched filtering analyses, and a higher probability that they are observed in the presence of non-Gaussian noise. We therefore present a method to search for gravitational waves from compact binary inspirals in next-generation detectors that is computationally efficient and robust against gaps in data collection and noise non-stationarities. Our method, based on the Hough Transform, finds tracks in the time/frequency plane of the detector that uniquely describe specific inspiraling systems. We find that we could detect $\sim 5$ overlapping, intermediate-strength signals (matched-filter signal-to-noise ratio $\rho\approx 58$) without a sensitivity loss. Additionally, we demonstrate that our method can enable multi-messenger astronomy: using only low frequencies ($2-20$ Hz), we could warn astronomers $\sim 2.5$ hours before a GW170817-like merger at 40 Mpc and provide a sky localization of $\sim 20$ deg$^2$ using only one ``L'' of Einstein Telescope. Additionally, assuming that primordial black holes exist, we derive projected constraints on the fraction of dark matter they could compose, $f_{\rm PBH}\sim 10^{-6}-10^{-4}$, for $\sim 1-0.1M_\odot$ equal-mass systems, respectively, using a rate suppression factor $f_{\rm sup}=2.5\times 10^{-3}$. Comparing matched filtering searches to our proposed method at a fixed sensitivity, we find a factor of $\sim10-$\fivehzimprove speed-up when we begin an analysis at a frequency of 5 Hz up to 12 Hz for a system with a chirp mass between $\mathcal{M}\in[1,2]M_\odot$.

\end{abstract}

\maketitle

\section{Introduction}

The LIGO, Virgo, and KAGRA detectors have observed $\mathcal{O}(100)$ binary black hole, binary neutron star, and black hole/neutron star systems since 2015 \cite{KAGRA:2021vkt,Olsen:2022pin}. These achievements were made possible by extremely sensitive interferometers \cite{TheLIGOScientific:2014jea,TheVirgo:2014hva,KAGRA:2020tym}, and by extensive and computationally heavy searches over a wide range of masses and spins of these systems. Some of the major methods that have successfully detected these systems, e.g. PyCBC \cite{Allen:2004gu,Allen:2005fk,DalCanton:2014hxh,Usman:2015kfa,Nitz:2017svb,Davies:2020tsx} and gstlal \cite{Messick:2016aqy,Sachdev:2019vvd,Hanna:2019ezx,Cannon:2020qnf}, are based on matched filtering, the optimal signal processing technique that correlates a deterministic signal waveform with noisy data and looks for a match. The immense sensitivity of matched filtering, however, comes at a high computational cost, since each individual waveform, that is, each choice of masses, spins, etc., must be convolved with the data over every possible arrival time, over months of observation \cite{maggiore2008gravitational}.

% Furthermore, current-generation \gwh observatories have been sensitive to the merger portion of the life of binary black hole \cite{}, which requires extensive numerical relativity simulations to produce waveforms for this extremely relativistic process \cite{}.
% \cite{Bandopadhyay:2022tbi}
Currently, matched filtering searches have primarily focused on systems above a solar mass \cite{Venumadhav:2019lyq,Zackay:2019btq,KAGRA:2021vkt}, with some sub-solar searches being performed, although with restrictions on the parameter space, due to high computational costs of convolving long-duration waveforms over months of data \cite{Phukon:2021cus,Nitz:2021mzz,LIGOScientific:2022hai,LIGOScientific:2021job,Nitz:2022ltl}. In third-generation \gwh detectors, however, the low-frequency sensitivity will greatly improve \cite{Punturo:2010zz,Hild:2010id,Branchesi:2023mws, Reitze:2019iox,Evans:2021gyd,Gupta:2023lga}, which means that we will be more sensitive to the inspiral portion of all systems, and can therefore see a longer-duration signal than currently possible. In other words, such \gwh signals will spend much more time at low frequencies than their current-generation counterparts spend in the detector sensitive frequency band. This implies that the number of templates necessary to cover the search parameter space will also increase as the minimum searchable frequency decreases, since phase mismatches between the template and signal accumulate with signal duration \cite{Nitz:2021mzz}. In fact, the number of templates needed to cover the parameter space between $1M_\odot$ and $3M_\odot$ is very sensitive to the low-frequency cutoff: starting at 40 Hz, in current detectors, the number of templates needed is $\sim 4000$, while at 2 Hz, in \et, $\sim 150000$ templates are needed, almost two orders of magnitude more (and we have not even considered sub-solar mass objects) \cite{Bosi:2010glx}. To combat this problem, a method for ``hierarchical matched filtering'' has been proposed to alleviate some of the cost, showing reductions in computational cost by a factor of a few to an order of magnitude, depending on the signal-to-noise ratio threshold without sensitivity loss \cite{Dhurkunde:2021csz}. Furthermore, another hierarchical matched filtering method has shown comparable sensitivities for matched filtering analyses using current-generation \gwh detector data, speeding up a matched-filtering analysis by a factor of 20, while maintaining the same sensitivity as the original matched filtering search \cite{Soni:2021vls}, and has recently been improved to better estimate outlier significance \cite{Soni:2023veu}. Nonetheless, these algorithms have not yet been tested in the context of the overlapping and long-lived signal regimes; therefore, further investigation into their efficacy and computational cost are necessary as well. 

Though matched filtering has a large computational cost, it has been extensively used in \gwh searches and has been shown to work in the presence of occasional glitches and noise non-stationarities \cite{Usman:2015kfa, gw170817FIRST,Zackay:2019kkv,Mozzon:2020gwa,DalCanton:2020vpm, Kumar:2022tto}. However, matched filtering on real data has primarily focused on signals of short durations (up to $\sim 100$ s), in which the noise is relatively stationary, Gaussian apart from isolated glitches, and devoid of gaps.
But, when the signals last for longer in the detector, each of these tenets will no longer hold true for most signals \cite{Davis:2018yrz} --- a signal \emph{not} polluted by a glitch or another disturbance will likely be the exception, not the rule. 

In particular, the non-Gaussian nature of noise becomes relevant when estimating the noise power spectral density; the data are more likely to contain disturbances, e.g. \emph{lines} or \emph{glitches}; and the detector could turn off with or without warning, causing gaps, where, on either side of the gaps, the noise properties could differ \cite{LIGOScientific:2019hgc}. Furthermore, glitch subtraction algorithms may leave behind residuals for certain types of glitches, e.g. Koi fish ones \cite{Bondarescu:2023jcx}, which would compound if many glitches were present and subsequently subtracted out when a signal also appeared in the detector. Therefore, the aforementioned matched filtering methods should be tested in realistic cases for future detectors as well, and may benefit from glitch subtraction mechanisms used in \cwh searches \cite{Steltner:2021qjy}. And in general, all types of analyses will have to grow to handle these particular problems, which may be amplified in the future when multiple glitches appear during a signal's duration.

% Though matched filtering has a large computational cost, it has been extensively used in \gwh searches, partially because it has primarily been used to look for signals of short durations (up to $\sim 100$ s), in which the noise is relatively stationary, Gaussian apart from isolated glitches, and devoid of gaps. However, matched filtering can work to detect \gws in the presence of occasional glitches \cite{Usman:2015kfa, gw170817FIRST,DalCanton:2020vpm}, and some different methods and statistics have shown promising results to further mitigate the effects of the noise, e.g. \cite{Zackay:2019kkv,Mozzon:2020gwa,Kumar:2022tto}. 

An additional complication is the sheer number of detectable compact binaries in third-generation \gwh detectors. Current rate estimates predict the detection of {$\mathcal{O}(10^{5}-10^{6})$} binary black hole mergers per {year}, and {$\sim 7\times 10^4$} binary neutron star inspirals per {year} \cite{Maggiore:2019uih}, and estimates of detectable black hole and neutron star mergers are around {$\mathcal{O}(100)$} per {day} and {$\mathcal{O}$(few)} per {hour}, respectively \cite{Maggiore:2019uih}. The impact of overlapping signals on matched filtering and Bayesian parameter estimation has been recently studied, concluding that {significant biases may exist if two binary black hole systems coalesce within 0.5 seconds of each other \cite{Pizzati:2021apa}; or, in other study, if $\Mc$ is within $10^{-4}M_\odot$ or $10^{-1}M_\odot$ for \bnsh or \bbhh mergers, respectively, within 10 ms 
 of each other \cite{Himemoto:2021ukb}. Additionally, a recent study has shown that matched filter redshift reach could be reduced by between 8\%-40\% for the Einstein Telescope/Cosmic Explorer detectors in the presence of a ``confusion noise'' of overlapping signals \cite{Wu:2022pyg}, though these effects could be mitigated if the so-called ``null stream'' can be perfectly constructed, or by subtracting binary neutron star signals if one knows their exact parameters well enough. But, there are limitations to the effectiveness of the null stream in practice, e.g. having equally sensitive detectors in the triangle, or ensuring no residuals of the subtracted signals are leftover in the data \cite{Goncharov:2022dgl}. 
Many others have also investigated parameter estimation of overlapping signals \cite{Relton:2021cax,Antonelli:2021vwg,Smith:2021bqc,Janquart:2022nyz,Langendorff:2022fzq,Relton:2022whr,Alvey:2023naa}. However, in these cases, the authors only considered two signals overlapping at once, which may be a simpler situation than what will be present in the future, and have employed the whole frequency band, which would not be possible for early-warning alerts. Additionally, while mock data challenges have also been conducted, and shown that overlapping signals can be recovered, computational cost remains a problem below 10 Hz, and signal separation may be problematic at times far from the merger, since the signals will be closer in frequency than at the time of merger \cite{Regimbau:2012ir,Meacher:2015rex}. 

An additional advantage of next-generation \gwh detectors is the prospect to detect inspiraling systems well before they merge, allowing time for astronomers to scope out the possible sky positions for various electromagnetic counterparts. Early warning of a merger of a binary neutron star system would permit electromagnetic observations of the entire post-merger phase \cite{Kasen:2014toa,Metzger:2018uni}. At the moment, matched-filtering analyses and triangulation could allow sky localization of $\sim 1-100$ square degrees depending on whether \et or \ce are used exclusively, or together, for a binary consisting of $1.4M_\odot$ neutron stars \cite{Sachdev:2020lfd,Nitz:2021pbr,Banerjee:2022gkv}, and much progress has been made localizing \gwh signals in future detectors via the Fisher Matrix formalism and \cite{Zhao:2017cbb,Chan:2018csa,Iacovelli:2022bbs} and Bayesian inference \cite{Baral:2023xst}. However, the localization is heavily dependent on how far away from coalescence the merger is, ranging from $\mathcal{O}(100)$ square degrees hours before the merger to under $\mathcal{O}(1)$ square degree in the milliseconds before the merger \cite{Cannon:2011vi,Hu:2023hos}. Even now, some matched filtering analyses attempt to warn astronomers by considering a fraction of the total bandwidth of the signal, i.e.  from 10 Hz to $\sim 30-50$ Hz, but this could result in, at best, only a minute of warning time before a merger \cite{Sachdev:2020lfd}, motivating the need to go down to lower frequencies, where matched filtering analyses begin to incur much larger computational costs.

Thus, it is of immense interest to improve the sky localization of compact binary systems well before the merger, and develop methods robust against data quality issues and overlapping signals with reasonable computational costs. Semi-coherent \cwh methods that divide data taken over $\Tobs$ into coherent chunks of length $\TFFT$ that are combined incoherently, could aid in this effort. Though used in a different context -- that is,
in all-sky or directed searches for persistent, quasi-periodic, \gws from asymmetrically rotating, stable and newborn neutron stars \cite{Riles:2017evm,Tenorio:2021wmz,Piccinni:2022vsd,Riles:2022wwz,Miller:2023qph} from ultralight boson clouds around black holes \cite{DAntonio:2018sff, Isi:2018pzk,Sun:2019mqb,LIGOScientific:2021jlr}, from \pbh inspirals \cite{Miller:2020kmv,Miller:2021knj,Guo:2022sdd} and from \dm that could couple to the interferometers \cite{Pierce:2018xmy,Guo:2019ker,Miller:2020vsl,Michimura:2020vxn, LIGOScientific:2021odm,Vermeulen:2021epa,Miller:2022wxu,Miller:2023kkd} -- these methods have been extensively developed to handle gaps and noise non-stationarities, and could also handle the ``astrophysical'' problem of having too many sources. Additionally, the computational cost of these methods do not increase steeply as in matched filtering when the minimum searchable frequency decreases \cite{Owen:1995tm}. In the case of long-lived signals, the use of semi-coherent methods significantly reduces the computational cost: computational cost scales as $\Tobs^6$ for fully coherent, wide-parameter searches, but only as $\Tobs^2\TFFT^4$ for semi-coherent ones \cite{Prix2009}. In essence, semi-coherent searches uses coarser grids in the parameter space, thus allowing a significant reduction in computational cost with only a modest reduction in sensitivity that is compensated by using lower thresholds to avoid missing candidates -- see Sec. \ref{subsec:mfcomp} for the application of this principle to our case. Additionally, see \cite{Sieniawska:2019hmd,Tenorio:2021wmz,Piccinni:2022vsd,Riles:2022wwz,Miller:2023qyw} for recent reviews on computationally efficient semi-cohernet \cwh searches.

This paper is meant as a proof-of-concept study to show the utility of \cwh methods to detect inspiraling compact binaries in next-generation \gwh detectors, using a particular one, the Generalized Frequency-Hough \cite{Miller:2018rbg,Miller:2020kmv,Guo:2022sdd}, as an example. Here, we focus on the \gws emitted in the low-frequency portion of the inspiral of a compact binary system, i.e. from $2-20$ Hz. At such low frequencies, the signal will (1) be less dominated by relativistic effects, (2) spend significantly more time in that band than at higher frequencies, allowing for the steady accumulation of signal-to-noise ratio \cite{VelcaniThesis}, and (3) be well-localized for particular inspiraling systems, even for a single interferometer (meaning, only one ``L'' in the final ET configuration) \cite{maggiore2008gravitational,Astone:2014esa}. Methods for ``long-duration bursts'', i.e. of $\sim 500$ s, could also be used to detect inspiraling systems \cite{Tiwari:2015bda,KAGRA:2021bhs}; therefore, our \cwh methods could complement canonical matched filtering and long-duration burst analyses to detect and localize the source quickly and computationally efficiently. 

The outline of this paper is as follows: we describe the signal morphology to which we are sensitive, and the basics of the method, in Sec. \ref{sec:gwinsp} and Sec. \ref{sec:method}, respectively. We quantify the sensitivity of our method at different minimum frequencies and compared to the matched filter, and describe the parameter space to which we are sensitive, the robustness of our method against noise non-stationarities, and the ability to separate signals, in Sec. \ref{sec:sensitivity}. We quantity the sky localization possible with our method, and the amount of time available to warn astronomers of coalescing systems, in Sec. \ref{sec:ew}. We then project constraints on \bnsh and \pbh rates and abundances in Sec. \ref{sec:astro}, and conclude with prospects for future work in Sec. \ref{sec:concl}.

\section{Gravitational Waves from inspiraling compact objects}\label{sec:gwinsp}

\subsection{The signal}

The inspiral of two compact objects, many orbits away from the innermost stable circular orbit, can be approximated as two point masses in a circular orbit around their center of mass (see section 4.1 of \cite{maggiore2008gravitational}), whose orbital frequency $\omega_{\rm orb}$ is given by Kepler's law. When accounting for the loss of orbital energy due to gravitational-wave emission, the distance between the two compact objects decreases, which means that   $\omega_{\rm orb}$ increases. Equating the power lost due to gravitational-wave emission with the rate of change of the orbital energy of the system, and knowing that the gravitational-wave frequency $f_{\rm gw}=\pi \omega_{\rm orb}$,
we arrive at \cite{maggiore2008gravitational}:

\begin{equation}
    \dot{f}_{\rm gw}=\frac{96}{5}\pi^{8/3}\left(\frac{G\mathcal{M}}{c^3}\right)^{5/3} f_{\rm gw}^{11/3},
    \label{eqn:fdot_chirp}
\end{equation}
where $\dot{f}_{\rm gw}$ is the rate of change of the frequency (the spin-up), $\mathcal{M}\equiv\frac{(m_1m_2)^{3/5}}{(m_1+m_2)^{1/5}}$ is the chirp mass of the system, $c$ is the speed of light, and $G$ is Newton's gravitational constant.

Eq. \ref{eqn:fdot_chirp} is a power law, with a braking index $n=11/3$ and a constant of proportionality $k$:
% \tcb{
\be
k\equiv\frac{96}{5}\pi^{8/3}\left(\frac{G\mathcal{M}}{c^3}\right)^{5/3}~.
\ee
% } 
This type of signal can be searched for with techniques developed to detect transient continuous waves lasting $\mathcal{O}$(hours-days) that could come from remnants of \bnsh mergers or supernova \cite{owen1998gravitational,mytidis2015sensitivity,Sarin:2018vsi,Miller:2018rbg,Oliver:2018dpt,Sun:2018owi,Banagiri:2019obu}. 

Integrating Eq. \ref{eqn:fdot_chirp}, we obtain the frequency evolution:
\begin{equation}
f_{\rm gw}(t)=f_0\left[1-\frac{8}{3}kf_0^{8/3}(t-t_0)\right]^{-\frac{3}{8}}~,
\label{eqn:powlaws}
\end{equation}
where $t_0$ is a reference time for the gravitational-wave frequency $f_0$ and $t-t_0=\tcoal$ is the time to merger. We also solve Eq. \ref{eqn:powlaws} for $t_{\rm merg}$: 

\begin{equation}
    % t_{\rm merg}=\frac{15}{768}\pi^{-8/3}\left(\frac{c^3}{G\mathcal{M}}\right)^{-5/3} (f-f_0)^{-8/3 }
    t_{\rm merg}=\frac{3}{8}\frac{f^{-8/3}-f_0^{-8/3 }}{k}.
\end{equation}

The amplitude evolution of the signal over time is \cite{maggiore2008gravitational}:

\begin{equation}
h_0(t)=\frac{4}{d}\left(\frac{G \mathcal{M}}{c^2}\right)^{5/3}\left(\frac{\pi f_{\rm gw}(t)}{c}\right)^{2/3},
\label{eqn:h0}
\end{equation}
where $d$ is the distance to the source. With Eqs. \ref{eqn:powlaws}-\ref{eqn:h0}, we can completely characterize, at zero post-Newtonian order, the inspiral of two compact objects.

\subsection{Signal duration}

Since \et will be built underground, it will have significantly better low-frequency sensitivity than the current \gwh detectors \cite{Punturo:2010zz}; thus, \bnsh inspirals will spend a lot of time in the sensitivity band of the detector relative to those in current generation detectors. 

In Fig. \ref{fig:fs_fdots_const_mc}, we plot the frequency evolution as a function of the inspiral time,  coloring how $\dot{f}$ also changes with time, for fixed chirp mass binaries: $\mathcal{M}=[0.4353, 0.8706, 1.3058, 1.7411]M_\odot$. We can see that at the lowest chirp mass, the signal could spend months at $\sim 2-3$ Hz, while at the highest chirp mass, it could spend hours there in that range. $\dot{f}$ increases rapidly as a function of frequency and time, which will affect the sensitivity of our proposed method towards inspiraling systems (see Sec. \ref{sec:method}).

\begin{figure}
    \centering
    \includegraphics[width=0.49\textwidth]{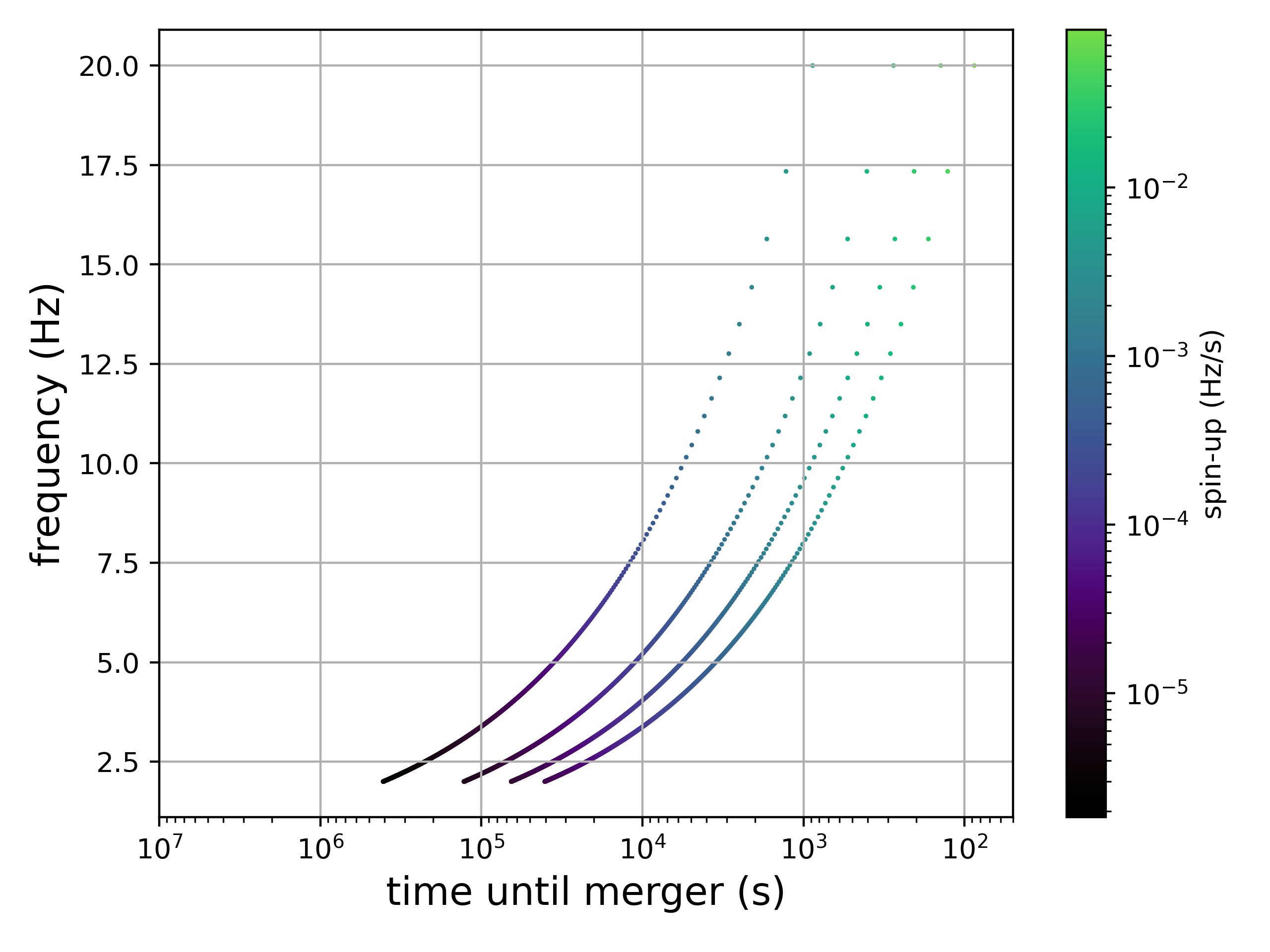}
    \caption{Chirp signals that show frequency evolution as a function of time to merger, with $\dot{f}$ colored. This figure indicates that we can observe for long periods of time at small $\dot{f}$ at the lower end of this frequency band. From left to right, each curve corresponds to: $\mathcal{M}=[0.4353, 0.8706, 1.3058, 1.7411]M_\odot$.}
    \label{fig:fs_fdots_const_mc}
\end{figure}

 In addition to Fig. \ref{fig:fs_fdots_const_mc}, we show, in Fig. \ref{fig:m1_m2_tobs}, the time that a \bnsh binary would spend in the $2-20$ Hz band, as a function of the individual component masses $m_1$ and $m_2$. 
Coupled with Fig. \ref{fig:fs_fdots_const_mc}, we see that signals with small $\dot{f}$ can last for a long time, making them similar to a \cwh signal. However, we will have to use a method that can handle not just quasi-monochromatic signals, but ones that follow a power law, as detailed in the next section.
\begin{figure}
    \centering
    \includegraphics[width=0.49\textwidth]{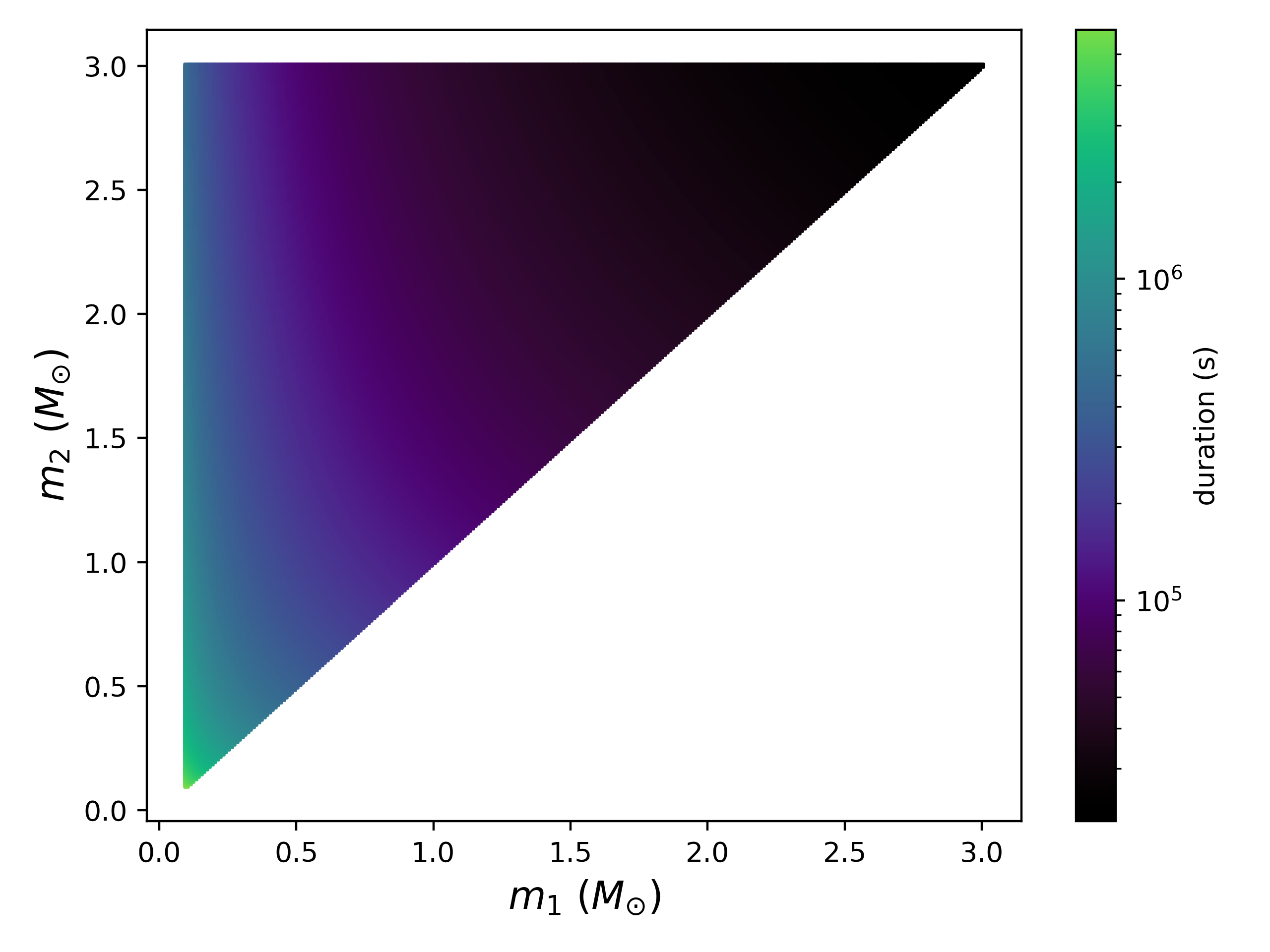}
    \caption{$m_1/m_2$ space describing how long signals would  last in the low-frequency ($2-20$ Hz) band.}
    \label{fig:m1_m2_tobs}
\end{figure}

\section{Search method: Generalized Frequency-Hough}\label{sec:method}

We propose a method based on the \GFH, a ``power-law track finder'', to search for inspiraling signals in \et and \ce. The \GFH transform is a pattern recognition technique that maps points in the time/frequency plane of the detector to lines in the frequency/chirp mass plane of the source, and was designed to search for signals that follow power-law frequency evolutions and that last $\mathcal{O}$(hours-days) \cite{Miller:2018rbg,Miller:2020kmv}. The method relies on making the following transformation of Eq. \ref{eqn:powlaws} to a new coordinate $x$:

\begin{equation}
x=f^{1-n}. \label{eqn:xx}
\end{equation}
Once we have changed coordinates (by substituting Eq. \ref{eqn:xx} into Eq. \ref{eqn:powlaws}), the signal's frequency evolution becomes linear in the new space:

\begin{equation}
x=x_0+k(n-1)(t-t_0),
\end{equation}
where we have also written $x_0=f_0^{1-n}$. Now, points in the time/$x$ plane are mapped to lines in the $x_0/k$ plane, and these two variables translate directly back to $f_0$ and $\mathcal{M}$. 

In contrast to matched-filtering searches, we perform the \GFH analysis on a time/frequency ``peakmap'', not a frequency series of Fourier transformed strain. We divide the strain $h(t)$ time-series into chunks of length $\TFFT$, fast Fourier transform each one (keeping the phase within each $\TFFT$), threshold the power in each frequency bin, and select local maxima above this threshold. The power in each time/frequency point that survives these two checks is called a ``peak'', and, for the purposes of the \GFH, we label each peak simply with a ``1'', and all other points 0. The \GFH, therefore, acts on a collection of 1s in the time/frequency plane - the value of the equalized power is not important.

While it should be better to sum raw power to obtain the highest possible sensitivity, the non-Gaussian, non-stationary nature of the noise allows strange artifacts to pop up throughout these chunks, which would effectively blind us to potential inspirals. This choice has been extensively studied in the context of \cwh Frequency-Hough searches \cite{Astone:2005fj,Astone:2014esa}, and has been shown to be robust against noise disturbances. In particular, powerful noise lines that appear at or wander around a certain frequency, or glitches that occur throughout the run, are only given a weight of ``1'' in the peakmap, thus greatly reducing their effects on real \gwh signals present in the data.

We choose $\TFFT$ on the basis of the spin-up, given in Eq. \ref{eqn:fdot_chirp}, by ensuring that the frequency modulation induced by $\dot{f}$ is confined to half a frequency bin, in each fast Fourier transform:

\begin{equation}
    \dot{f}\TFFT \leq \frac{1}{2\TFFT} \rightarrow \TFFT \leq \frac{1}{\sqrt{2\dot{f}}}.\label{eqn:tfft}
\end{equation}
The analysis choices that we make to construct the peakmap fix the sensitivity of the search, i.e. the choices of fast Fourier Transform length $\TFFT$ and duration of the map $\Tobs$.
The sensitivity of the \GFH search towards inspiraling binary systems has already been computed in \cite{Miller:2020vsl}, and is rewritten here, in terms of the maximum (luminosity) distance reach at a particular confidence level:

\begin{multline}
d_{\rm max}=\frac{C}{4}\left(\frac{G \mathcal{M}}{c^2}\right)^{5/3}\left(\frac{\pi}{c}\right)^{2/3} \frac{\TFFT}{\sqrt{\Tobs}}\left(\sum_i^N \frac{f^{4/3}_i}{S_n(f_i)}\right)^{1/2} \\ \times \left(\frac{p_0(1-p_0)}{Np^2_1}\right)^{-1/4}\sqrt{\frac{\theta_{\rm thr}}{\left(CR_{\rm thr}-\sqrt{2}\erfc^{-1}(2\Gamma)\right)}}.
\label{eqn:dmax}
\end{multline}
Here, $C$ is a geometric factor arising from averaging over an L-shaped ($=4.02$) or triangle-shaped ($=4.64$) detector,  $\theta_{\rm thr}$ is the threshold for peak selection selection in the equalized spectra when constructing the peakmap, $p_0=e^{-\theta_{\rm thr}}-$  $e^{-2\theta_{\rm thr}}$ $+\frac{1}{3}e^{-3\theta_{\rm thr}}$ is the probability of selecting a peak above the threshold $\theta_{\rm thr}$ if the data contains only noise, $p_1$ = $e^{-\theta_{\rm thr}}-$  2$e^{-2\theta_{\rm thr}}$ $+e^{-3\theta_{\rm thr}}$ , $CR_{\rm thr}$ is the threshold on the critical ratio we use to select candidates in the \GFH map, $N=\Tobs/\TFFT$, and $\Gamma$ is the chosen confidence level. 

\section{Sensitivity estimate}\label{sec:sensitivity}

\subsection{Optimized sensitivity}

Binary neutron star inspirals will be observable for a fraction of the total duration of an observing run in Einstein Telescope. In matched filtering, observing the signal for as long as possible, thereby accumulating signal power across all frequency bins, results in the best sensitivity towards a particular source, assuming that one can generate a waveform for the duration of the signal. However, in semi-coherent approaches, the need to break the data into chunks of $\TFFT$, based on the spin-up of the binary system, implies that it may actually be optimal to \emph{not} observe for the whole duration of the signal, but to cutoff the observation time at a fixed frequency. We currently do not vary $\TFFT$ as a function of the signal frequency, and typically pick $\TFFT$ based on the maximum $\dot{f}$ in a particular frequency band, which occurs at the highest frequency (Eq. \ref{eqn:fdot_chirp}). Therefore, if we observe for a shorter amount of time, i.e. not including the higher frequencies, $\dot{f}$ is smaller, meaning that $\TFFT$ can be longer. Thus, we consider the interplay between $\Tobs$ and $\TFFT$ as a function of the signal parameters over which we search using Eq. \ref{eqn:dmax}.
In Fig. \ref{fig:dist_vs_tfft}, we show the optimal distance reach as a function of $\TFFT$ length, which corresponds to a particular $\Tobs$ and bandwidth, shown in Fig. \ref{fig:dist_vs_tobs_bw}. The different curves refer to different chirp masses, as in Fig. \ref{fig:fs_fdots_const_mc}. Essentially, it does not pay to observe for as long as possible within a $2-20$ Hz band with the longest possible $\TFFT$; instead, we should cut off the observation time before the signal reaches 20 Hz, which corresponds to a fixed bandwidth given by the value on the colorbar in Fig. \ref{fig:dist_vs_tobs_bw}.

\begin{figure*}[ht!]
     \begin{center}
        \subfigure[ ]{%
            \label{fig:dist_vs_tfft}
            \includegraphics[width=0.5\textwidth]{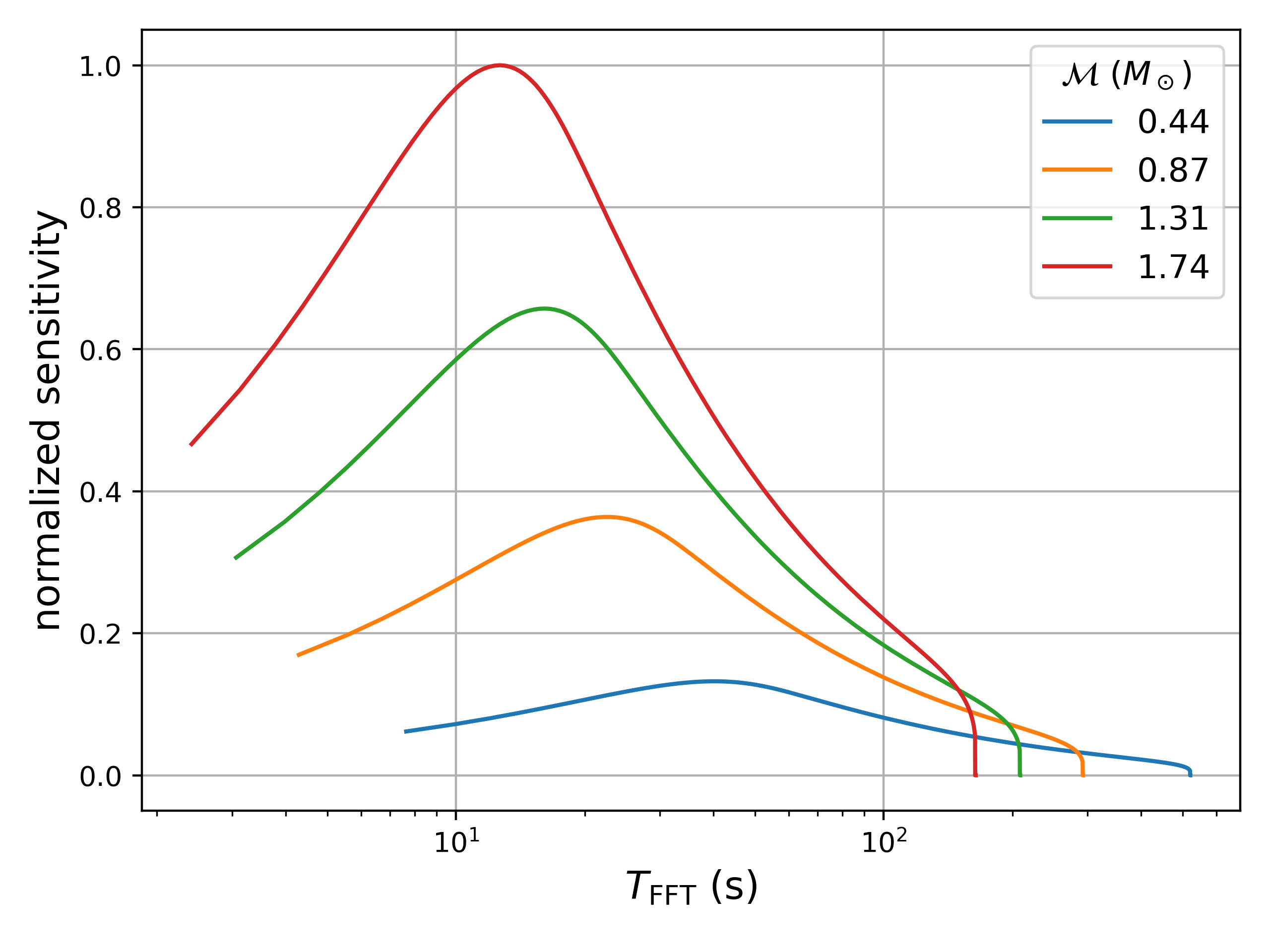}
        }%
        \subfigure[]{%
          \label{fig:dist_vs_tobs_bw}
           \includegraphics[width=0.5\textwidth]{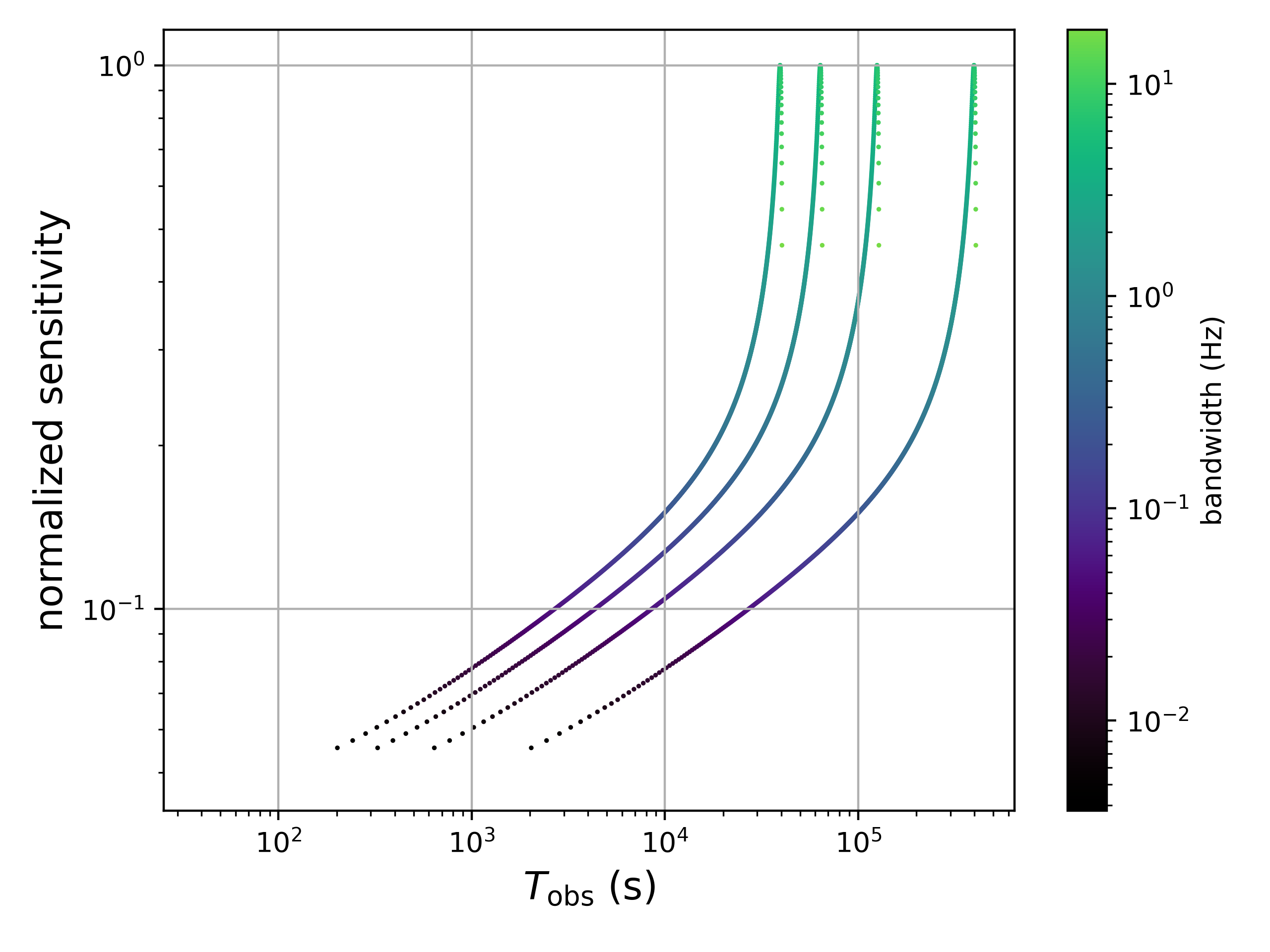}
        }\\ %  ------- End of the first row ----------------------%
    \end{center}
    \caption[]{%
   Left: normalized sensitivity as a function of $\TFFT$ length, for different chirp-mass systems, when analyzing $2-20$ Hz, when varying $\Tobs$ and $\TFFT$ in Eq. \ref{eqn:dmax}. The peak corresponds to the $\TFFT$ that maximizes our sensitivity to different compact binary coalescences. Right: Normalized sensitivity vs observation time, optimized by amount of frequency band analyzed and $\TFFT$, for a signal starting at 2 Hz. The chirp mass is indicated for each curve, and matches those in Fig. \ref{fig:fs_fdots_const_mc}.}%
     \label{fig:skymaps}
\end{figure*}

\subsection{Comparison to the ideal matched filter}
\label{subsec:mfcomp}
The matched filter can model exactly the inspiral portion of the \bnsh system. However, it comes at a very high computational cost.

We perform a quantitative comparison between the Generalized Frequency-Hough Transform and the matched filter. We borrow the formalism from \cite{Astone:2014esa} and derive the minimum possible amplitude that could be detected via matched filtering. 

The matched filter is defined as follows:

\begin{equation}
    \lambda \equiv 4\int_{f_{\rm min}}^{f_{\rm max}} df \frac{|\tilde{h}(f)|^2}{S_n(f)},
    \label{eqn:lambda}
\end{equation}
where $|\tilde{h}(f)|^2$ is the Fourier Transform of the inspiraling binary system, given by \cite{maggiore2008gravitational}

\begin{equation}
    |\tilde{h}(f)|^2 = \frac{5}{6}\frac{4}{25}\frac{1}{4\pi^{4/3}}\frac{c^2}{d^2}\left(\frac{G\mathcal{M}}{c^3}\right)^{5/3} f^{-7/3}.
    \label{eqn:hf2}
\end{equation}
Analogously to the Frequency-Hough \cite{Astone:2014esa}, we would like to compute the minimum detectable amplitude (or maximal distance reach) at a given confidence level for a matched filtering search, accounting for the fact that we have real search limitations, e.g. the need to select a fixed number of candidates. This information is encoded in a threshold on $\lambda$: since we know the distribution of spectral power is exponential, the probability that a particular frequency bin contains power $S$ larger than some threshold $S_{\rm thr}$ is:

\be
P(S>S_{\rm thr})=e^{-S_{\rm thr}}.
\ee
If we impose that the number of candidates above $S_{\rm thr}$ is $N_{\rm cand}$, then we
have $N_{\rm tot}\cdot e^{-S_{\rm thr}}=N_{\rm cand}$, so:
\be
S_{\rm thr}=-\log\left(\frac{N_{\rm cand}}{N_{\rm tot}}\right),
\ee
where $N_{\rm tot}$ is the total number of points in the source parameter space. This ratio is the false alarm probability of the search. If we fix $\frac{N_{\rm cand}}{N_{\rm tot}}=10^{-15}$, as in matched-filtering searches \cite{maggiore2008gravitational}, $S_{\rm thr}=34.54$. In practice, this ratio will be fixed based on the search that we actually perform, and the number of follow-ups that we can afford to do. In an all-sky search, this ratio is $\mathcal{O}(10^{-30})$ \cite{Astone:2014esa}, so our choice is quite conservative; however, we note that the sensitivity loss with respect to matched filtering does not differ by more than a factor of $\sim 2-3$  for even lower false alarm probabilities than what we choose.

The spectrum distribution in the presence of a signal of spectral amplitude $\lambda$ is a non-central $\chi ^2$ with
two degrees of freedom. The probability of having a spectrum value, in a given
frequency bin, larger than a threshold $S_{\rm thr}$ is then
\be
P(S>S_{\rm thr};\lambda)=\int_{S_{\rm thr}}^{\infty} dS
e^{-S-\frac{\lambda}{2}}I_0\left(\sqrt{2S\lambda}\right),
\label{eqn:detprob}
\ee
where $I_0$ is the zeroth-order Bessel function of the first kind. Eq. \ref{eqn:detprob} is the probability to detect a \gwh signal; therefore, if we would like to compute the minimum spectral amplitude that is detectable $95\%$ of the time, we set $P(S>S_{\rm thr};\lambda)=0.95$ and compute $\lambda_{\rm min}$ such that we achieve this probability by numerically integrating Eq. \ref{eqn:detprob}. We find $\lambda_{\rm min}=98$ . 

Now, combining equations \ref{eqn:h0} and \ref{eqn:hf2}, we can express $|\tilde{h}(f)|^2$ in terms of $h_0$:

\begin{equation}
    |\tilde{h}(f)|^2 = \frac{1}{30\pi^{8/3}}\frac{h_0^2}{16}\left(\frac{G\mathcal{M}}{c^3}\right)^{-5/3} f^{-11/3},
    \label{eqn:hf2_no-r}
\end{equation}
and then use Eq. \ref{eqn:lambda} to write $h_0$ in terms of $\lambda_{\rm min}$:
\begin{equation}
    h_{0,\rm min}^{\rm MF} = \sqrt{120\pi^{8/3}\lambda_{\rm min}\left(\frac{G\mathcal{M}}{c^3}\right)^{5/3} \left(\int_{f_{\rm min}}^{f_{\rm max}} df \frac{f^{-11/3}}{S_n(f)}\right)^{-1}}.
    \label{eqn:h0_mf}
\end{equation}
We can compare this expression to that computed for the semi-coherent \GFH search \cite{Miller:2020vsl}:

\begin{multline}
    h_{0,\rm min}^{\rm GFH}=\frac{4.02}{N^{1/4}\theta_{\rm thr}^{1/2}}\sqrt{\frac{N}{\TFFT}}\left(\sum_i^N \frac{f_i^{4/3}}{S_n(f_i)}\right)^{-1/2} f_0^{2/3} \\ \times \left(\frac{p_0(1-p_0)}{p^2_1}\right)^{1/4}\sqrt{\left(CR_{\rm thr}-\sqrt{2}\erfc^{-1}(2\Gamma)\right)},
    \label{eqn:h0_gfh}
\end{multline}
and compute the ratio

\begin{equation}
    R \equiv \frac{h_{0,\rm min}^{\rm GFH}}{h_{0,\rm min}^{\rm MF}}, \label{eqn:h0gfh_over_h0mf}
\end{equation}
in order to gauge how much worse our semi-coherent method will be, as a function of chirp mass. We show this ratio in Fig. \ref{fig:mfcomp}. We note that smaller chirp mass systems can be observed for longer times, but, as in \cite{Astone:2014esa}, this actually implies a greater sensitivity loss for our method compared to matched filtering. However,, the ``area of interest'' for binary neutron star inspirals is at least $\mathcal{O}(M_\odot)$, and here, the sensitivity loss is around $\sim 4$, and would be lower depending on our choice of false alarm probability, and the number of candidates that we could afford to follow up in a real search.

\begin{figure}
    \centering
    \includegraphics[width=0.49\textwidth]{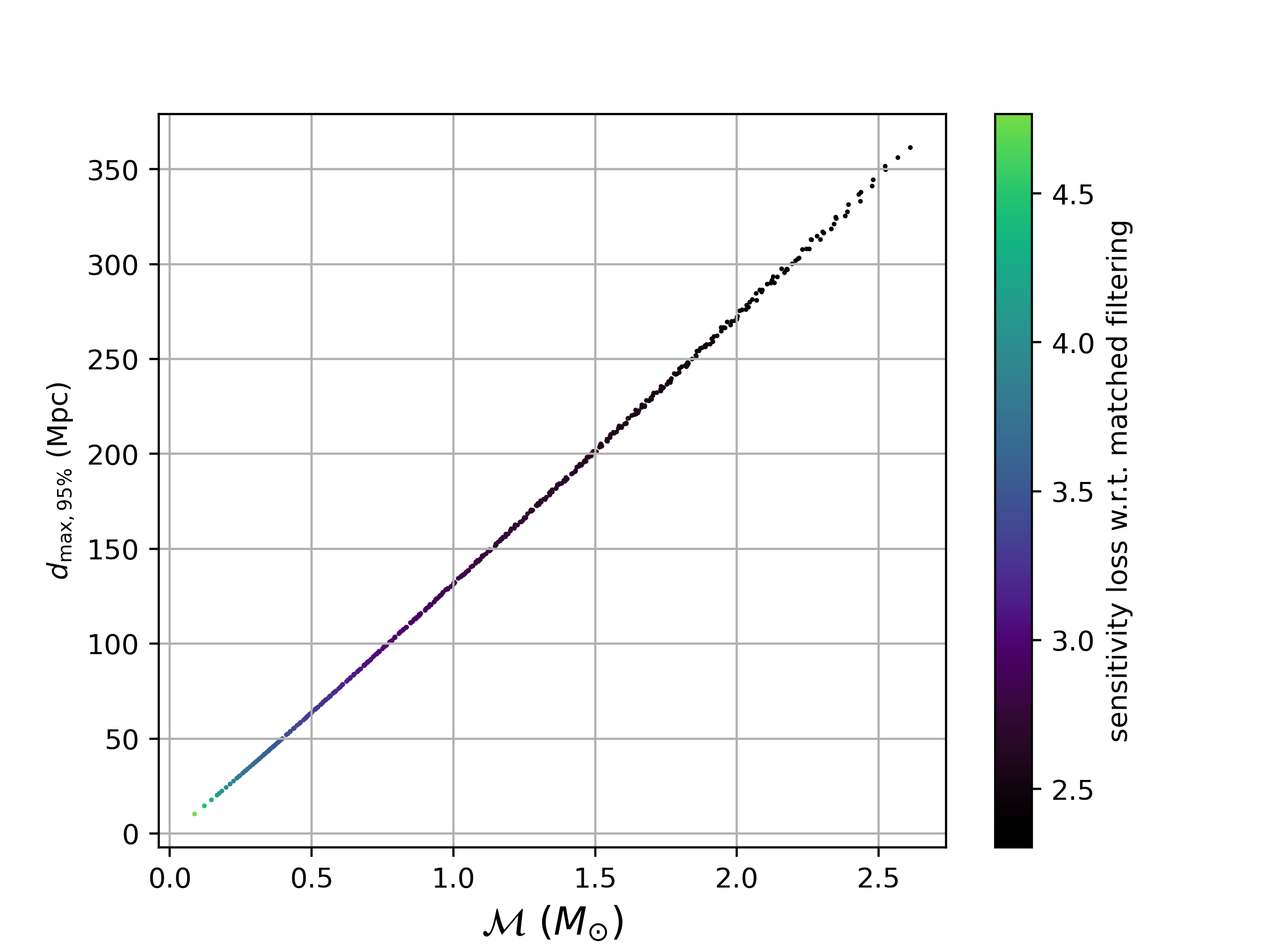}
    \caption{Luminosity distance reach and sensitivity loss with respect to matched filtering colored as a function of chirp mass. Observing for longer periods of time (corresponding to smaller chirp masses) increases the sensitivity loss compared to larger chirp masses, since matched filtering can accumulate more signal power over time compared to the semi-coherent \GFH, due to reducing the $\TFFT$ length (see Fig. \ref{fig:dist_vs_tfft}). Valid for signals that span $2-20$ Hz.}
    \label{fig:mfcomp}
\end{figure}

\subsection{Comparison to matched filtering searches}

While Fig. \ref{fig:mfcomp} indicates a sensitivity loss with respect when using our method versus the ideal matched filter on the same parameter space, we now consider practical matched filtering search configurations, and whether those will impact the relative sensitivity of our method.
In general, \cwh semi-coherent methods derive their utility from reduction in computational costs at the expense of sensitivity, as discussed in Sec. \ref{subsec:mfcomp}. However, we can compensate for this loss in sensitivity by, for example, setting a lower threshold on our detection statistic, in order to capture signals too weak to be completely detected in a ``coarse'' stage of the search \cite{Prix2009}, analogously to hierarchical matched filtering \cite{Soni:2021vls,Soni:2023veu}. 

In the case of binary inspirals in Einstein Telescope, lowering the minimum searched frequency as much as possible could allow our semi-coherent method to match or even exceed the sensitivity of matched filtering whose beginning frequency is greater than ours \cite{Bosi:2010glx}. We can therefore ask the question: \emph{at fixed sensitivity, how much more efficient will our semi-coherent method be compared to matched filtering?} The answer to this question will depend on detection thresholds and the chosen minimum frequency to analyze. 

For the purposes of this study, we fix a threshold on our semi-coherent method detection statistic of $CR_{\rm thr}=5$ \cite{Astone:2014esa,Branchesi:2023mws} and on the matched-filter \snr $\sqrt{\lambda}=\rho_{\rm thr} = 8$ \cite{maggiore2008gravitational}. Then, using Eq. \ref{eqn:h0_mf}, we calculate the minimum detectable amplitude of a matched filtering search $h_{0,\rm min}^{\rm MF} $ with different minimum frequencies $f_{\rm min,MF}$ ranging from 5 to 11 Hz, up to $\fmax=12$ Hz, for a system with $\mathcal{M}=2M_\odot$. Afterwards, with Eq. \ref{eqn:h0_gfh}, we compute the same minimum detectable amplitude $h_{0,\rm min}^{\rm GFH}$ for our semi-coherent search for the same range of frequencies for $f_{\rm min}$ using $\TFFT=14$ s, the ``optimal'' one. We then find the $f_{\rm min,MF}$ for which $h_{0,\rm min}^{\rm MF}=h_{0,\rm min}^{\rm GFH}$. To illustrate this procedure, in Fig. \ref{fig:et-same-sens} we plot $h_{0,\rm min}^{\rm MF}$ as a function of $f_{\rm min,MF}$, along with horizontal lines denoting $h_{0,\rm min}^{\rm GFH}$ at $f_{\rm min}=5\text{ and } 8$ Hz. The intersection of the horizontal lines with the curve determine at what $f_{\rm min,MF}$ the two searches would have equal sensitivities. In this case, the speed-up of using our semi-coherent method with respect to matched filtering is $\sim$\fivehzimprove and $\sim$\eighthzimprove starting at 5 and 8 Hz, respectively. 
Note that while we perform this comparison for $\mathcal{M}=2M_\odot$, the computational cost of the \GFH is in fact determined for range of chirp masses between $[1,3]M_\odot$ due to the grid constructed in $k$ ($\mathcal{M}$)-- see \cite{Miller:2018rbg,Miller:2020kmv} for more information on the grids. 

We can also ask the question: \emph{if we allow a sensitivity reduction of our method with respect to matched filtering, how  more efficient would our method be?} In Fig. \ref{fig:et-speed-up}, we plot the speed-up as a function of $f_{\rm min,MF}$, with the factor by which the strain sensitivity of our method is reduced with respect to matched filtering colored, for a system with $\mathcal{M}=2M_\odot$ and with $\fmin=5$ Hz and $\fmin=8$ Hz. We see that, if we are willing to accept a loss of $\sim 2$ in strain, the speed-up could be as much as a factor of $\sim$400. This enhancement in computational efficiency occurs because matched-filtering analyses are required to go to lower $f_{\rm min,MF}$ in order to be a factor of $\sim 2$ more sensitive than our method, which results in longer templates and a higher computational cost. Therefore, matched filtering analyses pay heavily in computing power to reach the lower frequencies that are necessary for early warning, while our method, though slightly less sensitive, is much more computationally efficient and would enable the use of low-frequency information at a fraction of the computational cost. See App. \ref{app:compcost} for details on how we compute the computational cost for each method.

\begin{figure*}[ht!]
     \begin{center}
        \subfigure[ ]{%
            \label{fig:et-same-sens}
            \includegraphics[width=0.5\textwidth]{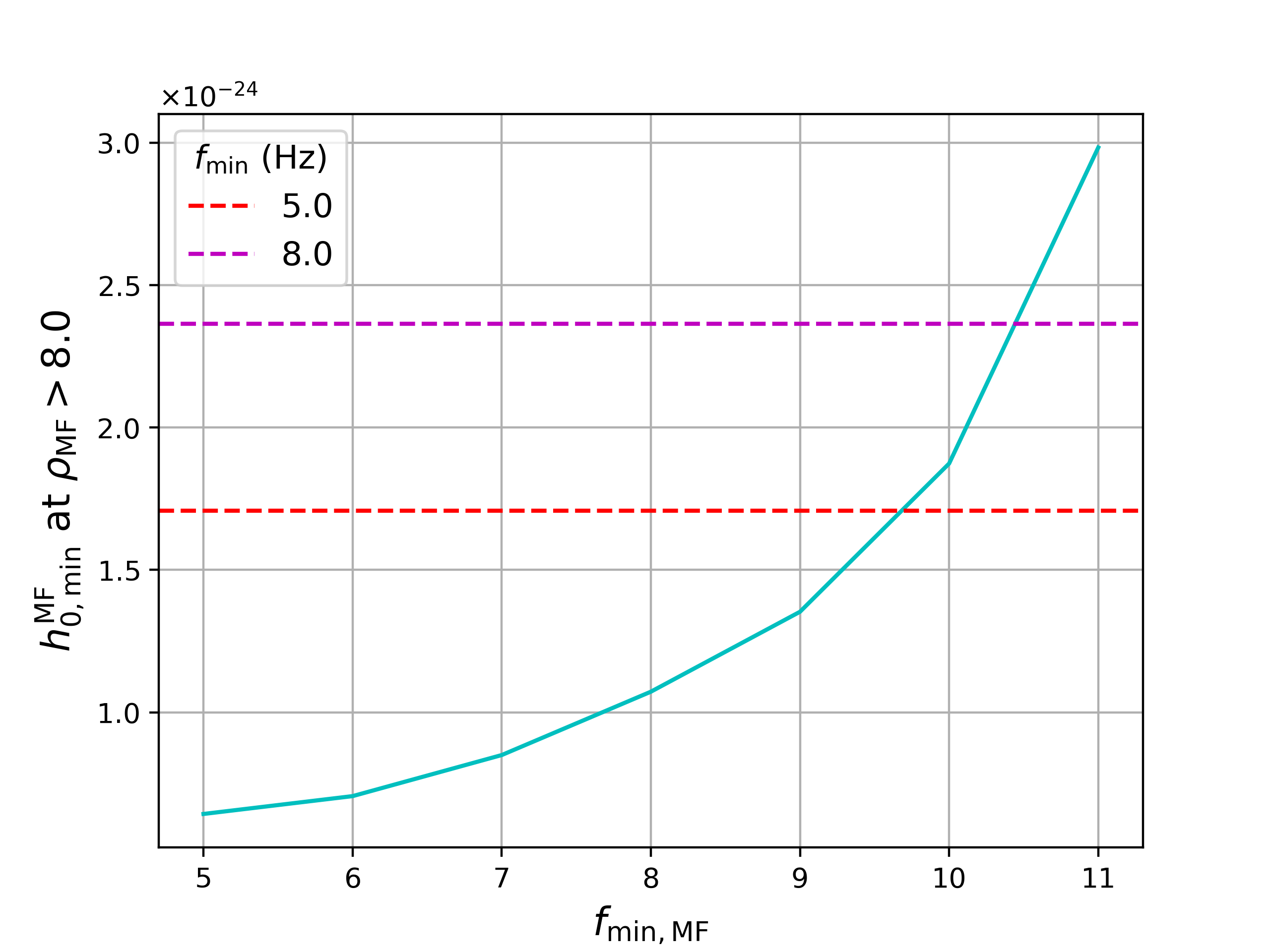}
        }%
        \subfigure[]{%
           \label{fig:et-speed-up}
           \includegraphics[width=0.5\textwidth]{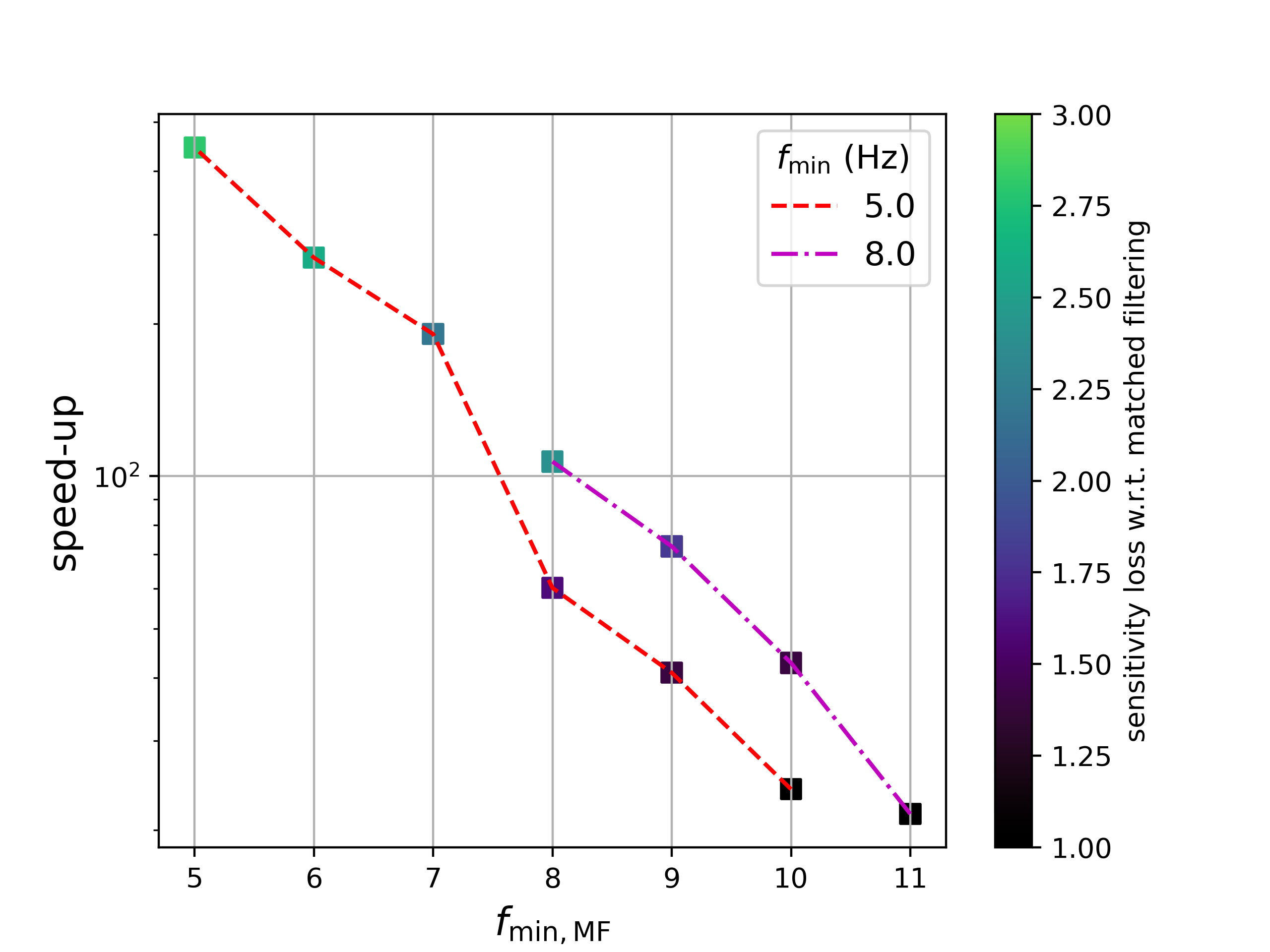}
        }\\ %  ------- End of the first row ----------------------%
    \end{center}
    \caption[]{Left: minimum detectable amplitude with matched filtering as a function of the frequency at which an analysis begins $f_{\rm min,MF}$. The colored dashed lines represent $h_{\rm 0,min}^{\rm GFH}$ computed using a threshold $CR_{\rm thr}=5$ when beginning an analysis of a $\mathcal{M}=2M_\odot$ system at 5 Hz (red) or 8 Hz (magenta). The intersection of these two horizontal lines with the curve represent when matched filtering and the \GFH have equal sensitivities. Right: the computational speed-up as a function of $f_{\rm min,MF}$ for an analysis using the \GFH starting at 5 Hz (red) and 8 Hz (magenta), with the factor of sensitivity loss in strain with respect to matched filtering colored.}%
     \label{fig:et-speed-up-same-sens}
\end{figure*}

We also would like to understand the impact on computational efficiency as a function of the chirp mass, at a fixed $\fmin=5$ Hz, which is shown in Fig. \ref{fig:et-speed-up-different-mcs}. Here, we can obtain speed-ups of at least $\sim 20$ for the lightest systems and $\sim 100$ for the heaviest ones without a loss of sensitivity, which increases to between $\sim 100-1000$ if we allow a sensitivity loss by a factor of $\sim 2-3$. We see that for lighter systems, at fixed sensitivity, the speed-up is smaller, since these signals would spend longer in the observing band, thereby allowing matched filtering to accumulate more \snr over its duration. This is consistent with \cite{Astone:2014esa}, Eq. 79, and our Eq. \ref{eqn:h0gfh_over_h0mf}, which both predict that over longer observation times, the sensitivity loss of semi-coherent to coherent methods increases.

\begin{figure}
    \centering
    \includegraphics[width=0.49\textwidth]{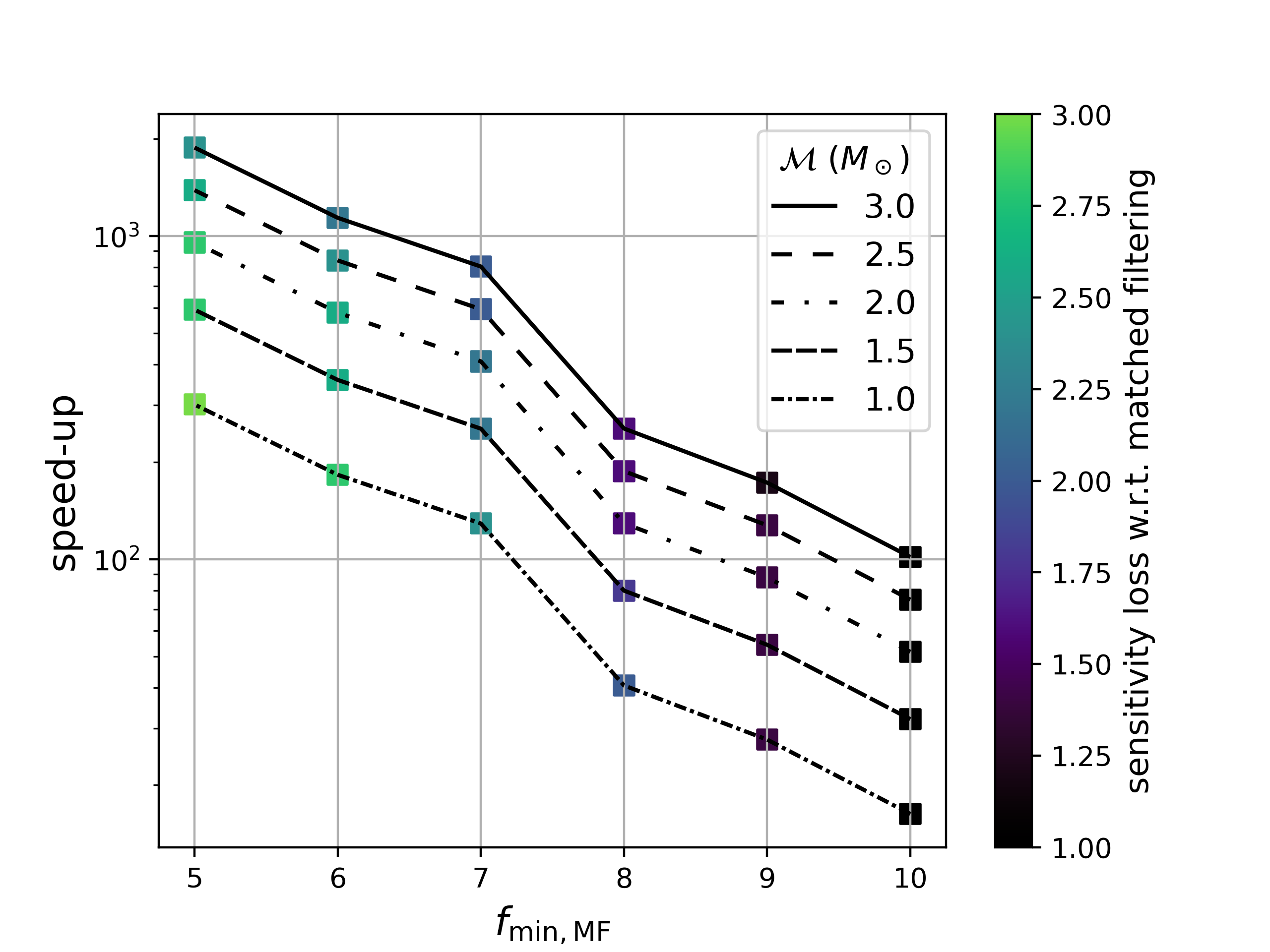}
\caption[]{The speed-up of our \GFH search with respect to matched filtering as a function of the minimum matched-filtering frequency and sensitivity loss, if we start analyzing at $\fmin=5$ Hz for systems with different chirp masses.}%
     \label{fig:et-speed-up-different-mcs}

\end{figure}

\subsection{Data gaps, non-stationary noise and overlapping signals}

The \GFH sums the presence of a peak in the time/frequency peakmap. Therefore, we \emph{inherently} work with data that are not continuous in the time and frequency domains, meaning that gaps, for any period of time, do not pose a systemic problem to our method. Furthermore, the detector power spectral density is estimated quickly in each FFT we take using an auto-regressive method \cite{Astone:2005fj}, meaning that changing, non-stationary noise, or different noise properties at one end of a gap and the other, do not affect our method's ability to work.

We illustrate these concepts in Fig. \ref{fig:pm}, in which we plot a time/frequency peakmap with ten signals that exist over the same time and frequency ranges in the left-hand panel (the white spaces are gaps due to applying a threshold on this map). The \GFH maps points in the peakmap to lines in the frequency/chirp mass plane of the source, and we can see here that each of the ten signals is well localized in a different pixel in Fig. \ref{fig:hm}. This occurs because each system has a different frequency at the start of the observation, and a different chirp mass. Of course, we could have also considered signals that have the same start frequency with different chirp masses, or the same chirp mass with different starting frequencies. However, the signals would still be localized into different pixels in Fig. \ref{fig:hm}. The power of the \GFH is that it sums time/frequency peaks along certain \emph{independent} tracks, ensuring that signal parameters are well localized in the frequency/chirp mass plane. 

We have also studied the impact of a large number of signals present in the data on the performance of the \GFH. In Fig. \ref{fig:eff-Ninj}, we plot the fraction of detectable signals as a function of the number of injections present in the peakmap $N_{\rm inj}$, marginalizing over starting frequencies uniformally distributed between [4.01,6.97] Hz, chirp masses  ([0.33,1.14]$M_\odot$), and signal durations ([200,10000] seconds). For each $N_{\rm inj}$, we performed 50 simulations, for three different signal amplitudes. We can see that while a small number, $\sim 5$, of simultaneously present number of injections does not impact the efficiency, also shown in \cite{lianysreport}, the efficiency degrades between $10-100$ injections. We note that this efficiency could be improved via a better estimation of the auto-regressive spectrum used to construct the peakmaps \cite{Pierini:2022wgc}, which has been optimized for singular, weak monochromatic signals. Moreover, our efficiency represents the realistic case in which we do not know which signals will be present in any given peakmap -- of course, if we only considered a couple of signals present simultaneously, and also constructed the peakmap with the appropriate $\TFFT$ to be optimally sensitive to each signal, these efficiencies would improve. However, when the detector turns on, we will not know which signals are present, so we cannot tune $\TFFT$ and the size of the peakmap for each signal. Therefore, our results represent a realistic test-case of unknown overlapping signals.

\begin{figure*}[ht!]
     \begin{center}
        \subfigure[ ]{%
            \label{fig:pm}
            \includegraphics[width=0.5\textwidth]{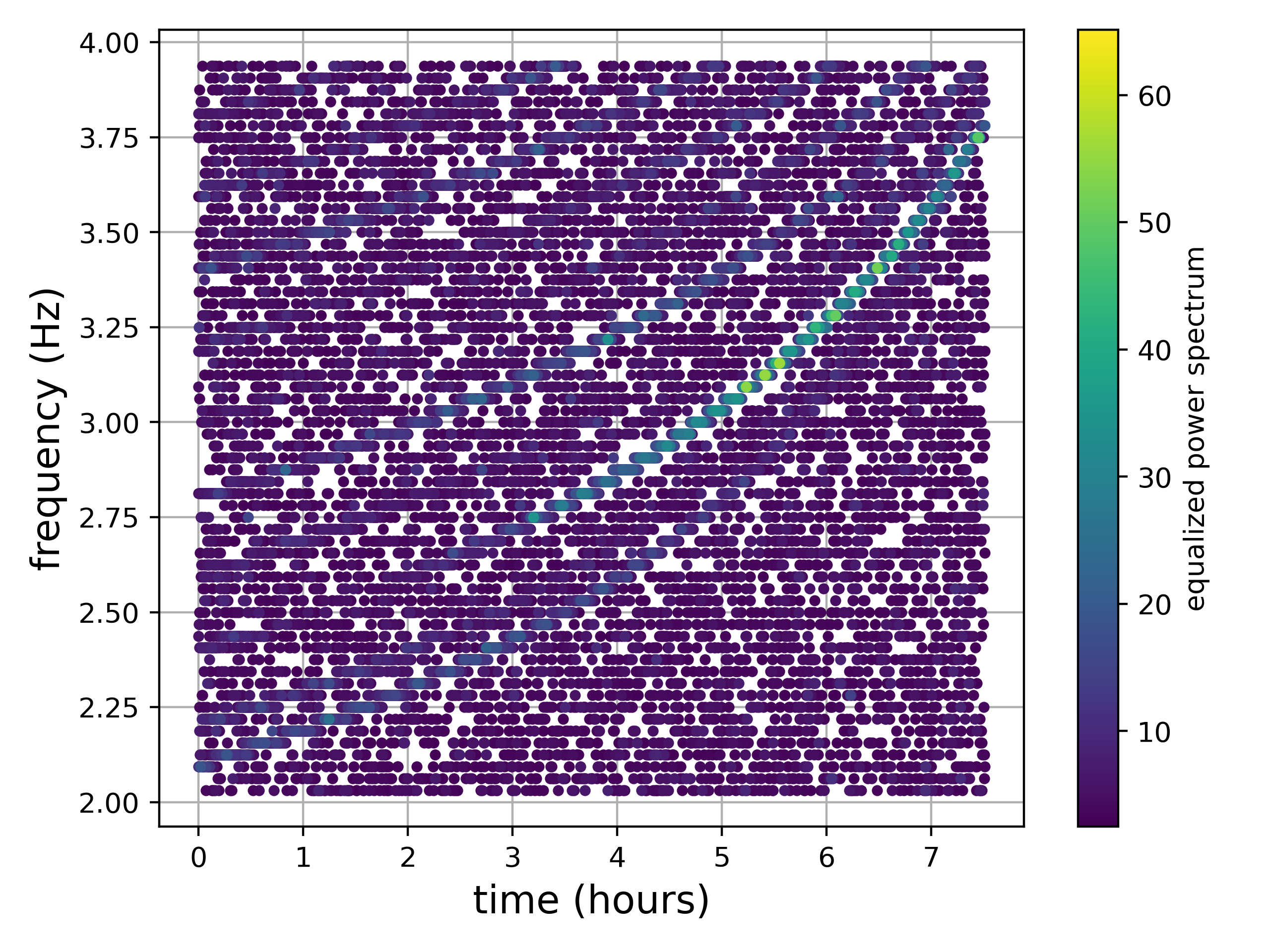}
        }%
        \subfigure[]{%
           \label{fig:hm}
           \includegraphics[width=0.5\textwidth]{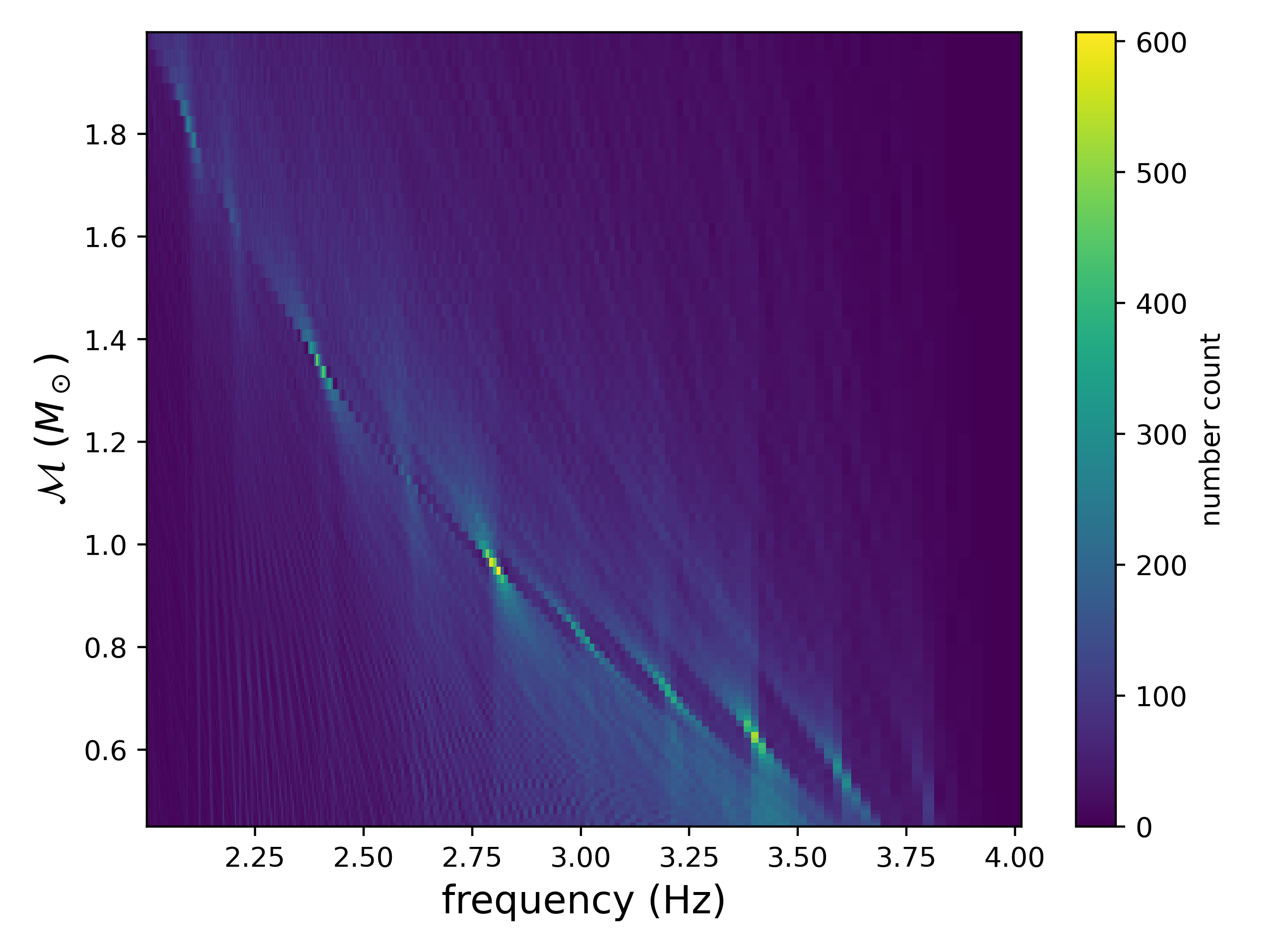}
        }\\ %  ------- End of the first row ----------------------%
    \end{center}
    \caption[]{%
Ten signals have been injected in white noise between 2-4 Hz, and are recovered by the Hough transform. Signals overlapping in time and in frequency can be easily isolated in the Hough plane. $\sqrt{S_n}\sim 8\times 10^{-24} \text{ Hz}^{-1/2}$. The amplitude for all signals is the same at $t=0$ ($h_0=10^{-23}$), but over time, each changes in different ways based on the frequency evolution and the chirp mass of the injected signal. The matched-filter signal-to-noise ratio $\rho\approx 58$. In the case of non-Gaussian, non-stationary noise, this method has been tested extensively in \cite{Miller:2018rbg,Miller:2019jtp,LIGOScientific:2018urg}.
     }%
     \label{fig:pm_hm_injs}
\end{figure*}

\begin{figure}
    \centering
    \includegraphics[width=0.49\textwidth]{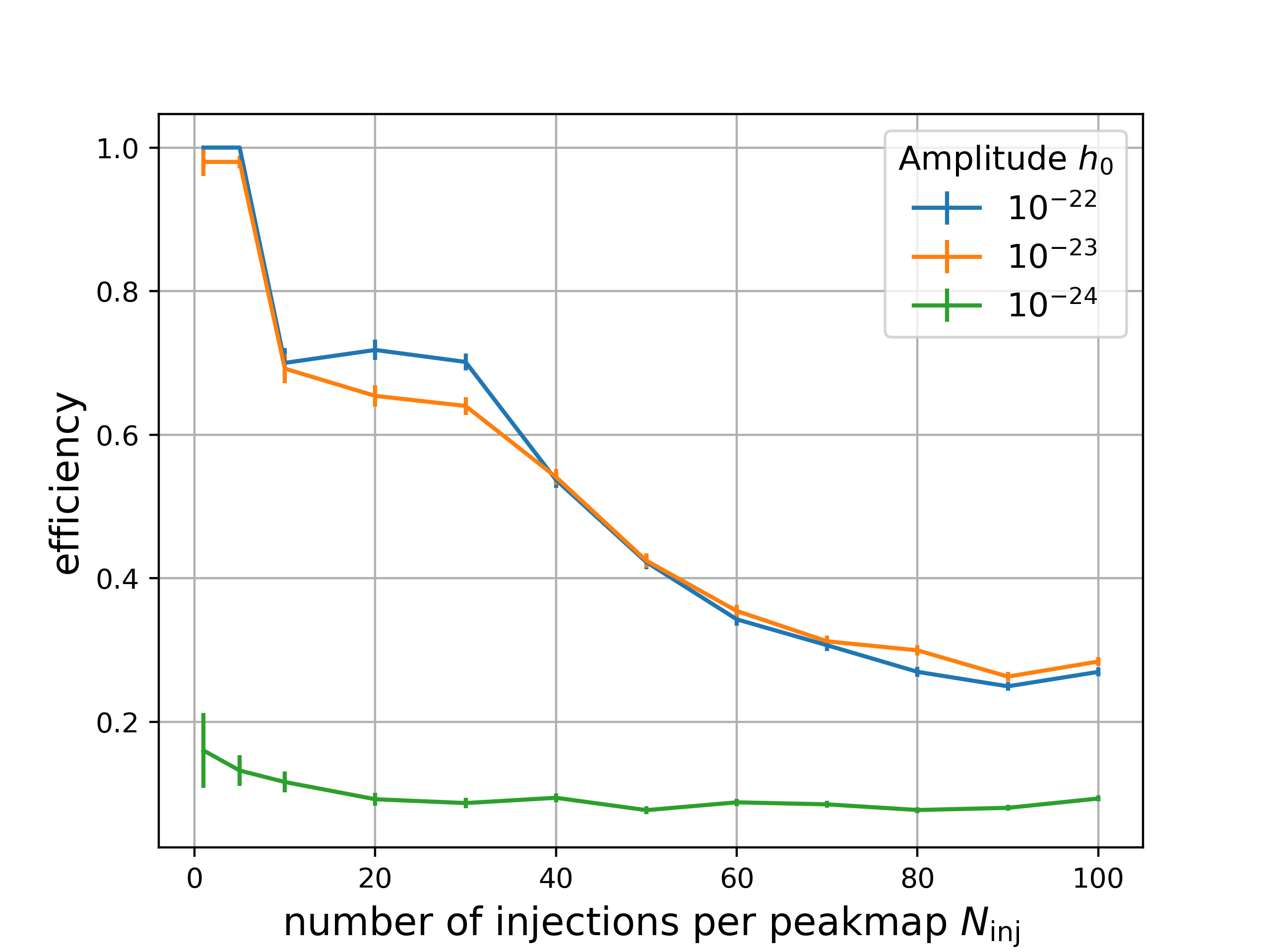}
    \caption{Efficiency as a function of number of injected signals per peakmap, where 50 simulations at each $N_{\rm inj}$ were performed. Even for strong signals, efficiency degrades rapidly past $N_{\rm inj}\sim 30$. ``Efficiency'' is defined as the fraction of the total number of signals injected across all simulations that are recovered by the \GFH.}
    \label{fig:eff-Ninj}
\end{figure}

\subsection{Sensitivity robustness against different $\fmin$}

Throughout this paper, we have consistently chosen the minimum frequency at which we compute sensitivity estimates to be $\fmin=2$ Hz. However, the true frequency floor will vary depending on the nature of the noise in Einstein Telescope, and it is not clear yet whether such amazing sensitivity at 2 Hz will be achievable \cite{Beker:2014ata,Badaracco:2020qmm}. We therefore consider how our results will change with respect to differing $\fmin$. We vary the beginning frequency $\fmin$ and compute the sensitivity, noting that a higher $\fmin$ implies shorter observation time and less accumulated power, as shown in Fig. \ref{fig:varyfmin}. Our sensitivity appears to vary by by no more than a factor of 10 across the chosen $\mathcal{M}$ with respect to the $\fmin=2$ Hz case. We note that in Fig. \ref{fig:varyfmin}, the curves for each chirp mass do not all extend to 10 Hz because we set a threshold of at least 10 minutes to observe, and higher chirp mass systems will not last for longer than that between $\fmin$ and 20 Hz.

\begin{figure}
    \centering
    \includegraphics[width=0.49\textwidth]{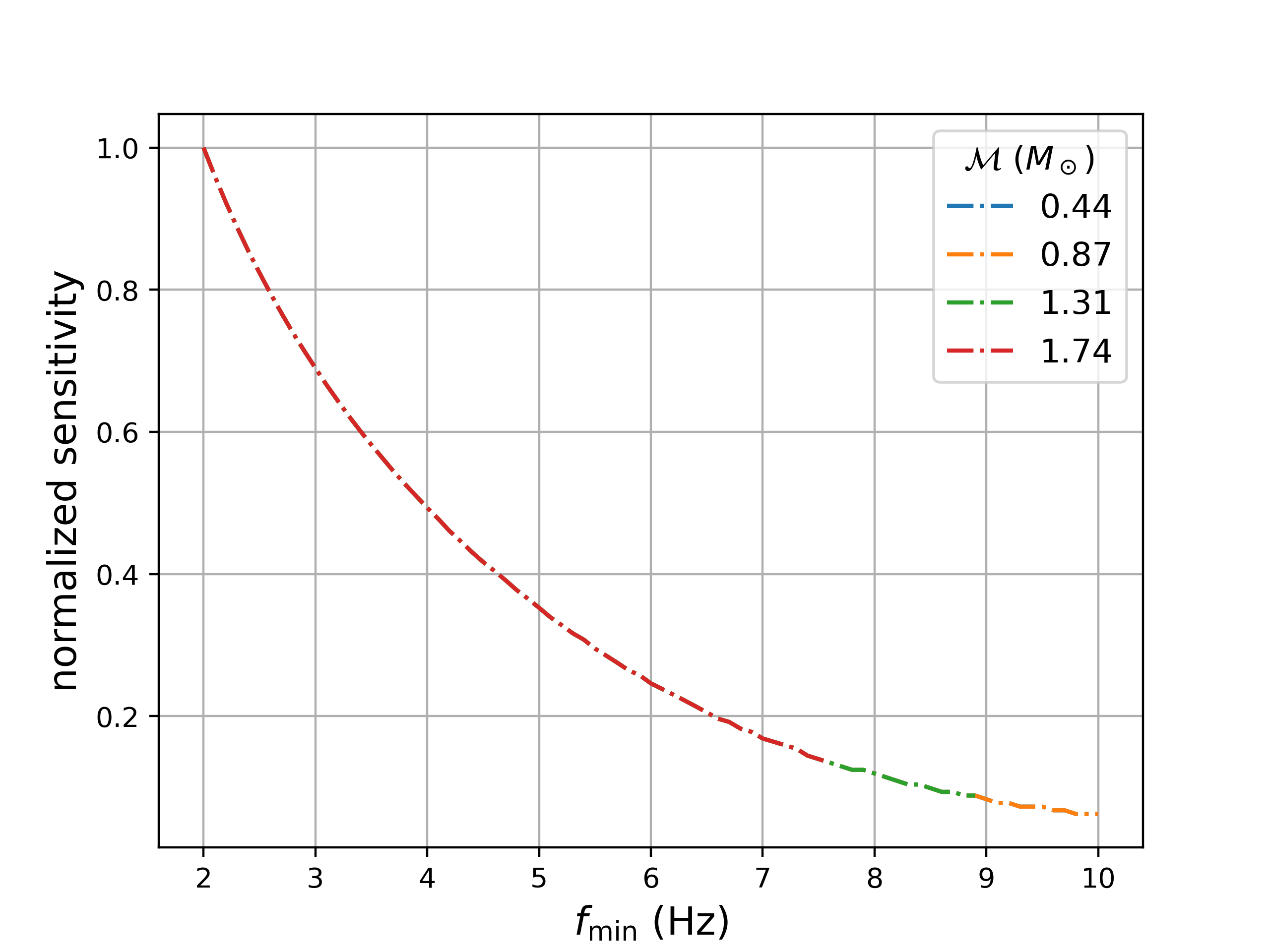}
    \caption{Normalized sensitivity as a function of the beginning analysis frequency. At $\fmin=5$ Hz, we see that we would only be 35\% as sensitive to inspiraling systems compared to if we observed them starting from $\fmin=2$ Hz.}
    \label{fig:varyfmin}
\end{figure}

\section{Enabling multi-messenger astronomy}\label{sec:ew}

\subsection{Sky localization}

Part of the need for early warning is to provide astronomers not only with intrinsic source parameters but also sky position. In our method, depending on the chirp mass of the system, the frequencies at which we observe it, and the $\TFFT$ with which we analyze the data, the accuracy of sky localization will be different.

\begin{figure}
    \centering
    \includegraphics[width=0.49\textwidth]{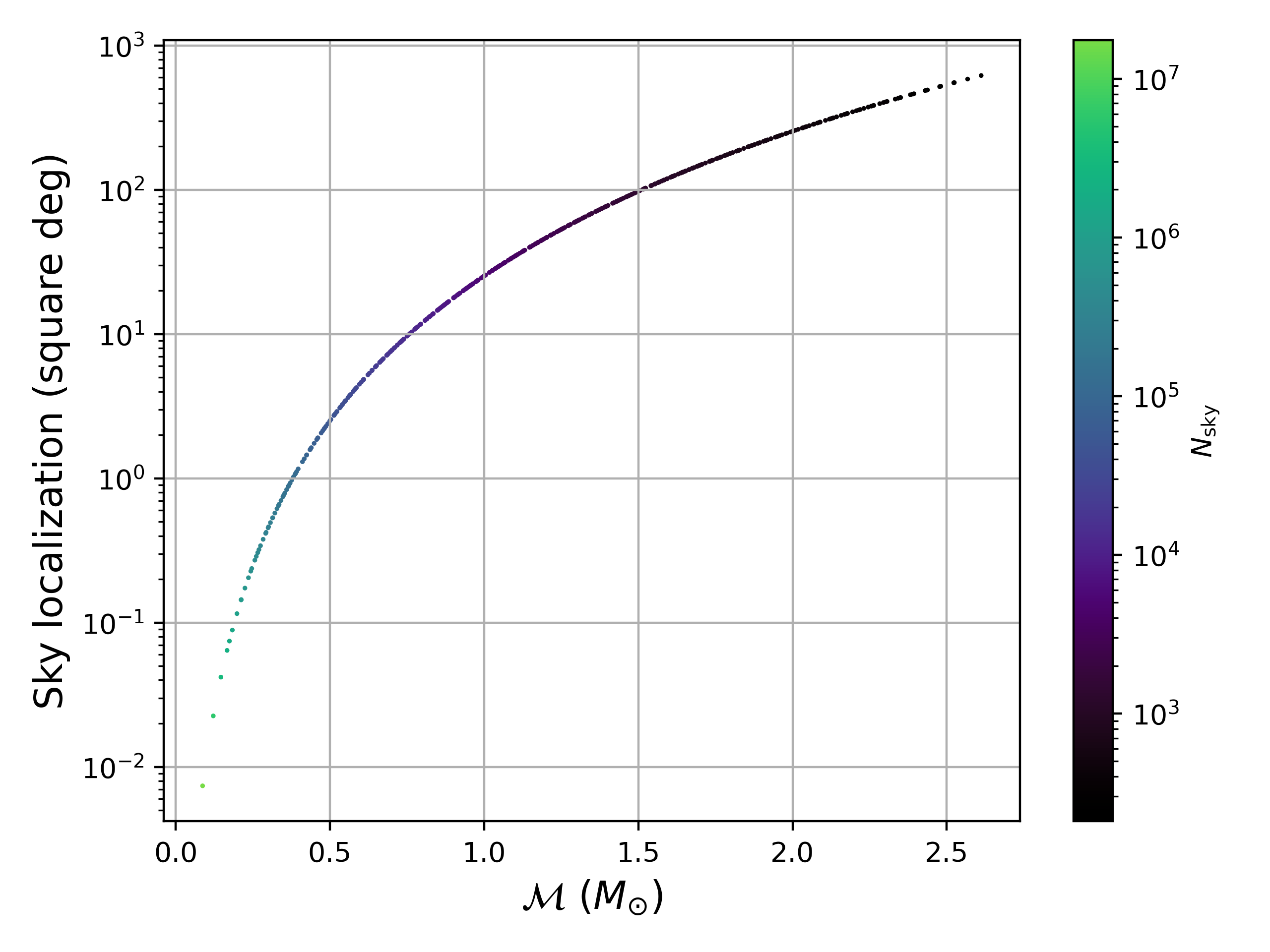}
    \caption{Finest possible sky localization based on a single-interferometer analysis as a function of chirp mass, assuming that the signal frequency evolution can be completely demodulated after estimating the chirp mass and initial frequency of the system, resulting in a purely monochromatic signal. For an equal-mass system of $1.4M_\odot$ ($\Mc=1.22M_\odot$), this plot is consistent with the estimations present in \cite{Nitz:2021pbr} for a single \et interferometer, though we consider only the $2-20$ Hz band.}
    \label{fig:optimal_skyloc}
\end{figure}

The \GFH will provide an estimation of orbital frequency and chirp mass. In an initial search, for high chirp masses $\mathcal{O}(1M_\odot)$, $\TFFT$ and the \gwh frequency are too small to construct a sky grid (that is to say, we cannot perform sky localization in the first pass of this method). This is because because the Doppler shift is directly proportional to the frequency, and since $\TFFT$ is quite small compared to its values \cwh all-sky searches, the Doppler shift is contained within one (large) frequency bin, since the increase in \gwh frequency due to the inspiral is greater than the frequency shift induced by the Doppler motion. However, after we estimate \gwh frequency and chirp mass, we can correct for the phase evolution of the signal, neglecting higher-order post-Newtonian corrections. 
If a perfect correction is made, the signal would become monochromatic; thus, we would be able to set $\TFFT\sim \Tobs$. 
In practice, we cannot make a perfect correction given the coarseness in the chirp mass and \gwh frequency grids, but we can understand how our resolution in the sky will improve with each pass of our method. As in standard all-sky searches, we would perform a \emph{hierarchical} follow-up, in which we will increase $\TFFT$ gradually, usually by a factor of 2 in each pass \cite{KAGRA:2022dwb}. As we constrain more and more the signal parameters, we will also obtain a finer and finer resolution in the sky, but of course, the time that will remain to warn astronomers will decrease. 
% The design of this follow-up method in detail, and its impact on early warning times, will be explored in future work.

Assuming that we can make a perfect correction and use $\TFFT=\Tobs$, in Fig. \ref{fig:optimal_skyloc}, we plot the finest possible sky resolution, in square degrees, and the number of sky points in the grid, as a function of chirp mass, using a single interferometer (one ``L'' of a triangular \et instrument). 
This figure is generated by following the formalism in \cite{Astone:2014esa},  Eq. 35$-$43, which accounts for the Doppler modulation when constructing a grid on the sky to perform a \cwh analysis. We can see that the sky localization for higher chirp mass systems is worse than that for weaker ones. This is due to the fact that the higher chirp mass systems have larger spin-ups, and thus have smaller durations. In order to achieve this localization, in practice we will need to progressively increase $\TFFT$ after recovering $f_0$ and $\Mc$, and allow for the uncertainties in each parameter.%, 

\begin{figure*}[ht!]
     \begin{center}
        \subfigure[ ]{\label{fig:time_ew}%
 \includegraphics[width=0.49\textwidth]{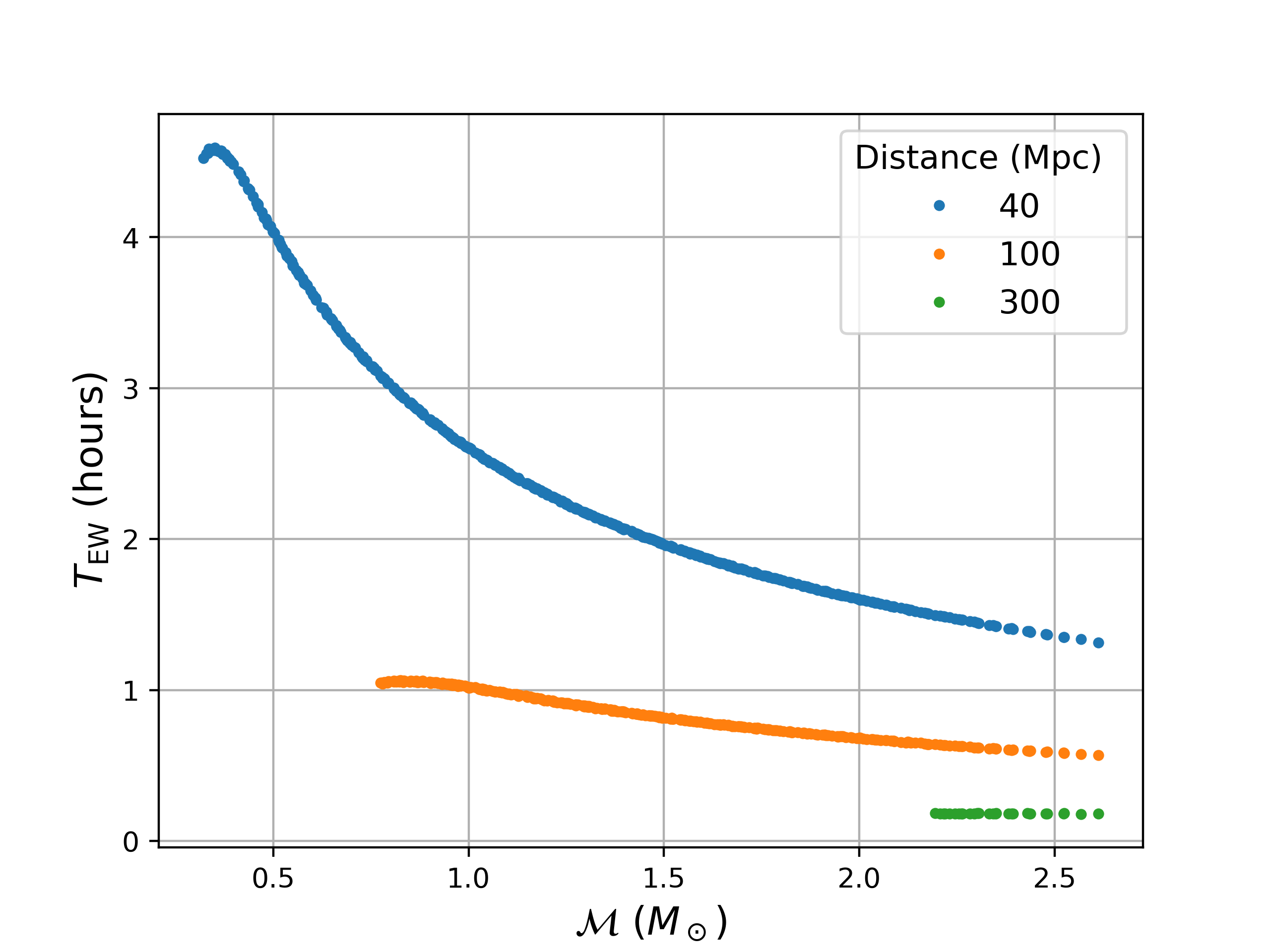}
        }%
        \subfigure[]{\label{fig:frac_tcoal}%
 \includegraphics[width=0.49\textwidth]{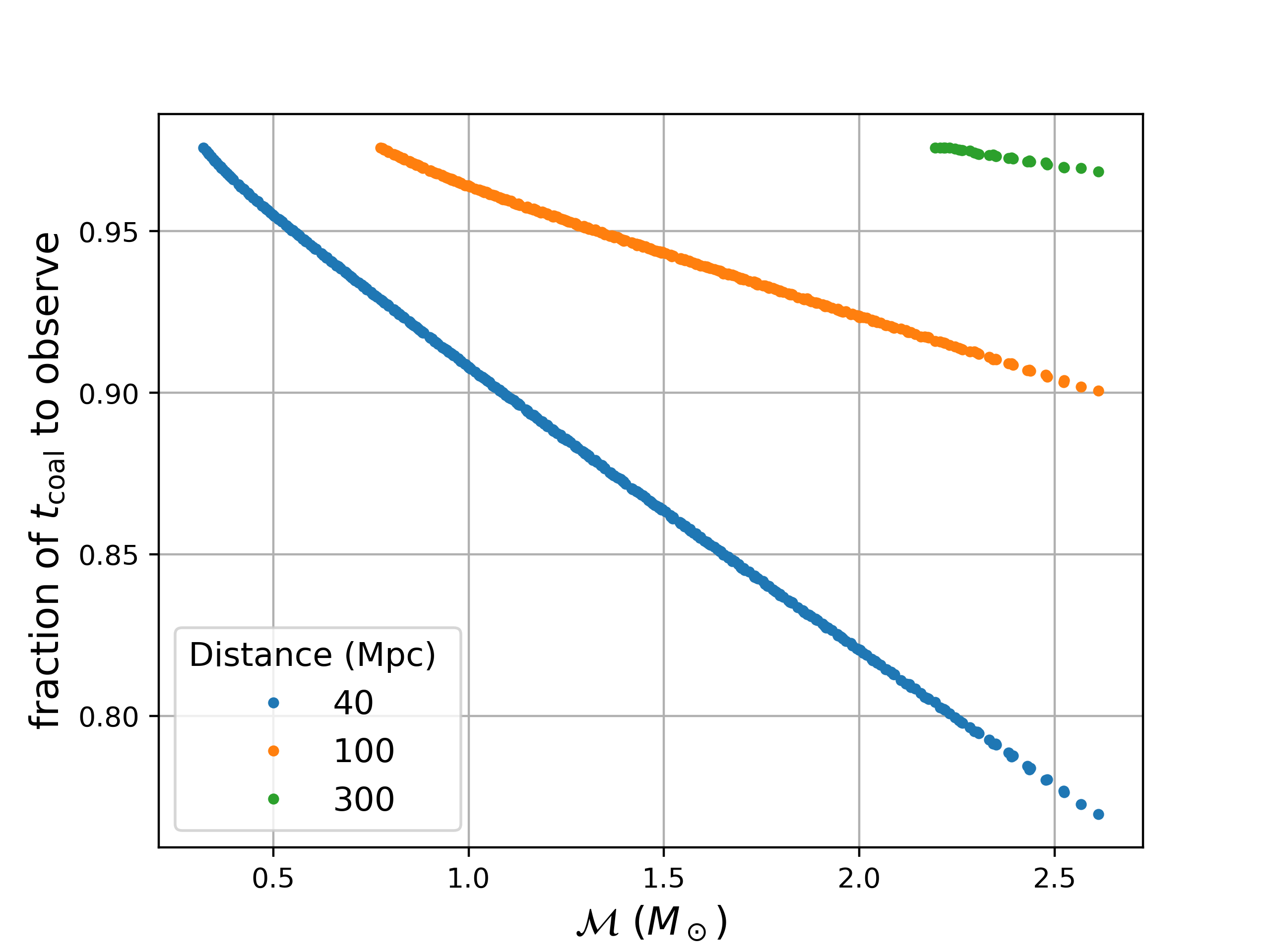}
        }\\ %  ------- End of the first row ----------------------%
    \end{center}
\caption{Left: Maximum time available to warn astronomers that a merger of two compact objects will occur at different distances. Right: fraction of coalescence time necessary to observe a signal at a particular chirp mass is colored. At higher distances (and lower masses), more time is needed to obtain a given distance reach. If the source is closer, then we do not have to observe for as long as if the source were farther, at a fixed chirp mass. }
    \label{fig:time_ew_tcoal}
\end{figure*}

It is worth emphasizing here that our sky localization capabilities arise from using data from a ``single'' interferometer, that is, from one  ``L'' in an ET configuration.   Analyzing data separately from each ``L'' allows us to perform coincidences in order to reduce the false alarm probability, but not to improve the sensitivity, in contrast to matched filtering. So, only using one ``L'' is a good approximation for the sensitivity and sky localization capabilities.

\subsection{Early warning}

The next generation of \gwh detectors will be sensitive at low-enough frequency such that, in principle, ample time will exist to warn astronomers that a merger of two compact objects will happen somewhere in the sky.
We define the maximum time that we will have to warn astronomers, $\tew$, as:

\begin{equation}
\tew \equiv \tcoal-\Tobs,
\end{equation}
where $\Tobs$ is the time that we observe the inspiral such that we obtain the maximum distance reach, as indicated in Fig. \ref{fig:dist_vs_tobs_bw}, and $\tcoal$ is the time to coalescence, calculated at $\fmin=2$ Hz, the starting frequency of the band analyzed.

In Fig. \ref{fig:time_ew}, we plot $\tew$ as a function of $\Mc$ if, at that particular $\Mc$, we could detect an inspiraling system at least 40, 100 or 300 Mpc away. For the 40 Mpc curve, $\tew$ time peaks at a chirp mass of $\sim0.25M_{\odot}$, then steadily falls off. The peak occurs because of the interplay between the accumulation of signal-to-noise ratio over time, and the duration of the signal. At higher chirp masses, the signals are shorter, and a smaller fraction of $\tcoal$ is necessary to reach the chosen distance.

For the 100 Mpc curve in Fig. \ref{fig:time_ew}, systems below 1$M_{\odot}$ cannot be reached at 100 Mpc. In this case, we are observing a little bit less and less as we increase the chirp mass, but we still need to observe for a large fraction of $\tcoal$ (Fig. \ref{fig:frac_tcoal}), $\mathcal{O}(10^{4})$ seconds for all $\Mc>1M_{\odot}$. We note that, as expected, $\tew$ is smaller for larger distances, since we need to observe for a larger fraction of $\tcoal$ to reach a larger distance.

\section{Projected merger rates and constraints}\label{sec:astro}

As with any method to search for inspiraling systems in next-generation detectors, we can provide estimates of how well we can constrain various astrophysical quantities in the future, including the merger rates of compact objects, as well as \dmh properties. We describe how we will obtain these constraints in the following two subsections.

\subsection{Neutron-star merger rate densities}

In order to compute merger rate densities for these systems, we adopt the formalism present in \cite{Biswas:2007ni}, recalling the following equation:

\begin{equation}
    \mathcal{R}_{95\%,i} = \frac{3.00}{\avgVT_i },
\end{equation}
where $\mathcal{R}_{95\%,i}$ is the merger rate density at a given chirp mass $i$, and $\avgVT_i$ is the average space-time volume enclosed for a given chirp mass, given by \cite{LIGOScientific:2016kwr,LIGOScientific:2016ebi,Singh:2021zah,Singh:2023cxn}:

\begin{equation}
  \label{eqn:average-space-time-volume}
  \avgVT_i = T \int \dd z \, \dd \theta \, \diff{V_c}{z} \frac{1}{1+z} s_i(\theta) f(z,\theta),
\end{equation}
where $f(z,\theta)=\Gamma$ in Eq. \ref{eqn:dmax} and, in this case, is calculated from inputting a variety of possible source luminosity distances, and $\theta=\Mc$. $\diff{V_c}{z}$ is the differential co-moving volume as a function of redshift, whose values depend only on cosmology, and is given in App. \ref{app:vt}.

We make the assumption as in \cite{Kim:2002uw}: that the population follows the
observed sources:
\begin{equation}
  \label{eqn:delta-population}
  s_i(\theta) = \delta\left( \theta - \theta_i \right),
\end{equation}
where $\delta$ is the Dirac delta function and $\theta_i$ are the parameters of source type $i$.

From these equations and the distance reaches computed in earlier sections, we can arrive at merger rate density estimates for compact binary systems with different chirp masses, given in Fig. \ref{fig:bnsrate}. These projected constraints can be used to exclude binary population evolution models as to provide crucial information about the evolutionary scenarios. We test these constraints against some of the binary evolution models for population I and II field BNSs, whose details are given in App. \ref{app:pop_synth}. We find that after $\sim 13$ years of observation, these models would begin to be excluded with our method.

\begin{figure}
    \centering
    \includegraphics[width=0.49\textwidth]{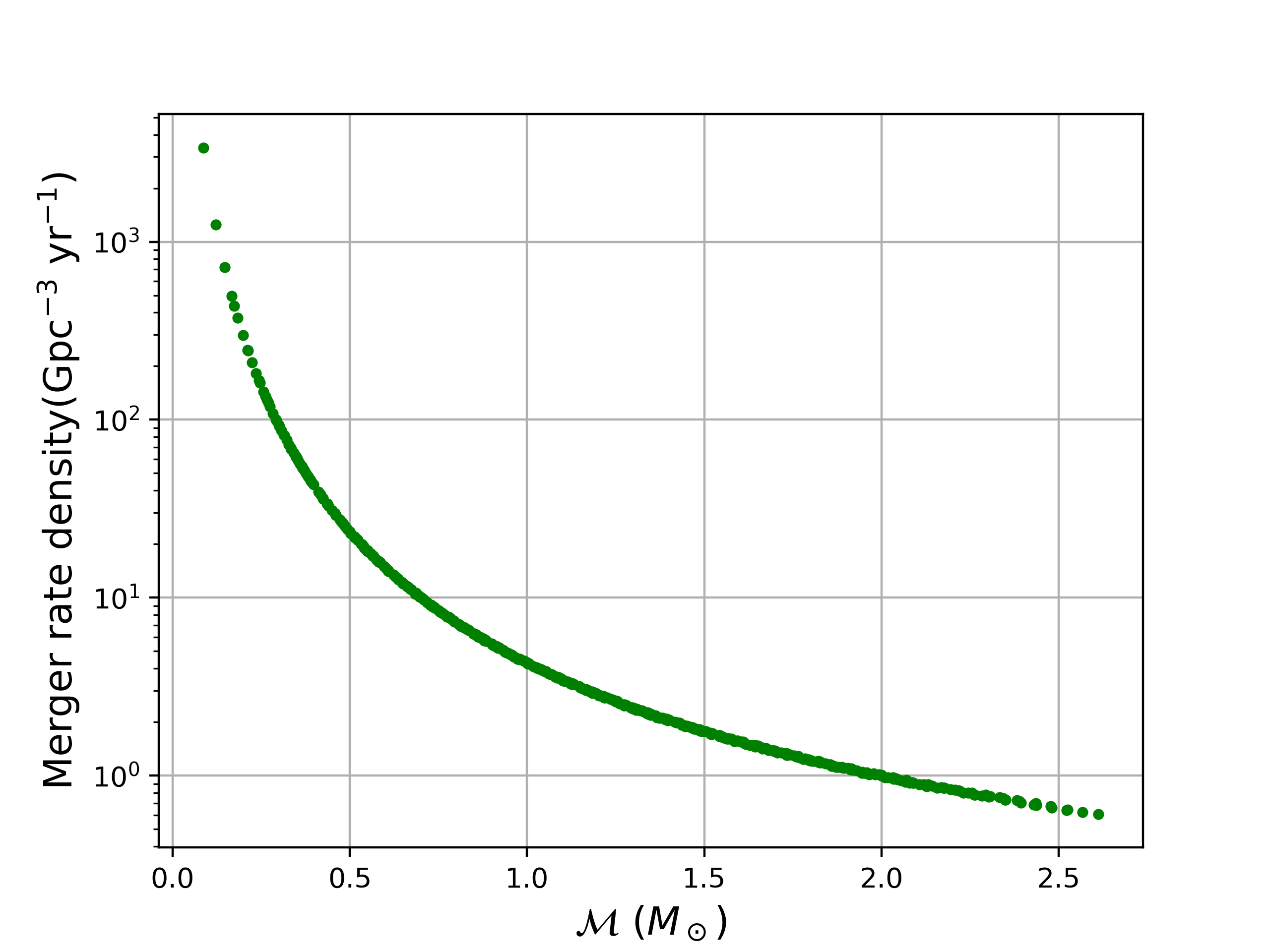}
    \caption{Projected merger rate densities, treating each chirp mass as a separate population of compact objects.}
    \label{fig:bnsrate}
\end{figure}

\subsection{Primordial black hole binaries}

While we have called the inspiraling compact objects in this paper ``neutron stars'', we could have easily named them ``\pbhs'', especially those with sub-solar masses \cite{Carr:2019kxo,Green:2020jor}. In fact, the \GFH only considers two objects with a certain chirp mass, and makes no assumptions about what these objects actually are.

By using cosmological rate predictions for early \pbh binaries and of \pbh binaries in clusters, the rate densities in Fig. \ref{fig:bnsrate} can be translated into projected constraints on the dark matter fraction of \pbhs. 
We use the relationships in ~\cite{raidal2019formation,Hutsi:2020sol} for the cosmological merger rates that assume a purely Poissonian \pbh spatial separation at formation, given by:

\begin{eqnarray}
\mathcal{R}_{95\%,i} =& 1.04 \times 10^{-6}\, \mathrm{kpc}^{-3} \mathrm{yr}^{-1} f_{\rm sup} f(m_{\rm PBH})^2 \nonumber\\
&\left(\frac{m_\mathrm{PBH}}{M_\odot}\right)^{-32/37}  \left(f_{\rm PBH}\right)^{53/37},
\label{eqn:rate-equal}
\end{eqnarray}
which correspond to the rate per unit of logarithmic mass of the two binary black hole components $m_{\rm PBH}$. $f_{\rm PBH}$ is the \dm fraction of \dm that \pbhs could constitute, and $f(m)$ is the mass distribution function of \pbhs normalized to one ($\int f(m) {\rm d} \ln m = 1$).  We have included a suppression factor $f_{\rm sup}$ that reduces the rates of \pbh formation due to gravitational influence of early forming \pbh clusters, nearby \pbhs and matter inhomogeneities~\cite{raidal2019formation}. 

For unequal-mass mergers, we consider the merging rate in the limit $m_1 \gg m_2$:
\begin{eqnarray}
\mathcal{R}_{95\%,i} =& 5.28 \times 10^{-7}\, \mathrm{kpc}^{-3} \mathrm{yr}^{-1} f_{\rm sup} f(m_1) f(m_2) \nonumber\\
&\left(\frac{m_1}{M_\odot}\right)^{-32/37} \left(\frac{m_2}{m_1}\right)^{-34/37} \left(f_{\rm PBH}\right)^{53/37}~.
\label{eqn:rate_asymm}
\end{eqnarray}
and place constraints on an effective, model-independent parameter $\tilde{f}$, due to the heavy model-dependence of \(f_{\rm sup}\):

\begin{equation}
\tilde f^{53/37} \equiv f_{\rm sup} f(m_1) f(m_2) f_{\rm PBH}^{53/37}\label{eqn:ftilde}
\end{equation}
We choose to constrain $\tilde{f}$ because of the uncertainty in the value of $f_{\rm sup}$. For equal-mass \pbhs and $f_{\rm PBH}=1$, $f_{\rm sup} \approx 2 \times 10^{-3}$ \cite{Hutsi:2020sol,Clesse:2020ghq,raidal2019formation}, but $f_{\rm sup}$ could take on different values depending on the mass functions and mass ratios, eccentricities and tidal fields of binary systems {\cite{Eroshenko:2016hmn,Cholis:2016kqi}}.

In Fig. \ref{fig:pbhconstr}, we plot the projected constraints on equal-mass and asymmetric-mass ratio binaries using the merger rates inferred in Fig. \ref{fig:bnsrate}. For equal-mass systems (blue curve, blue text), in order to obtain a constraint on $\fpbh$, we assume $\fsup=2\times 10^{-3}$ and a monochromatic mass function, while for asymmetric mass-ratio binaries (black curve, black text), we constrain the effective parameter $\tilde{f}_{\rm asymm}$ for $m_1=2.5M_\odot$, motivated by the QCD phase transition \cite{Byrnes:2018clq}, and by observations of stellar-mass black holes.

\begin{figure}
    \centering
    \includegraphics[width=0.49\textwidth]{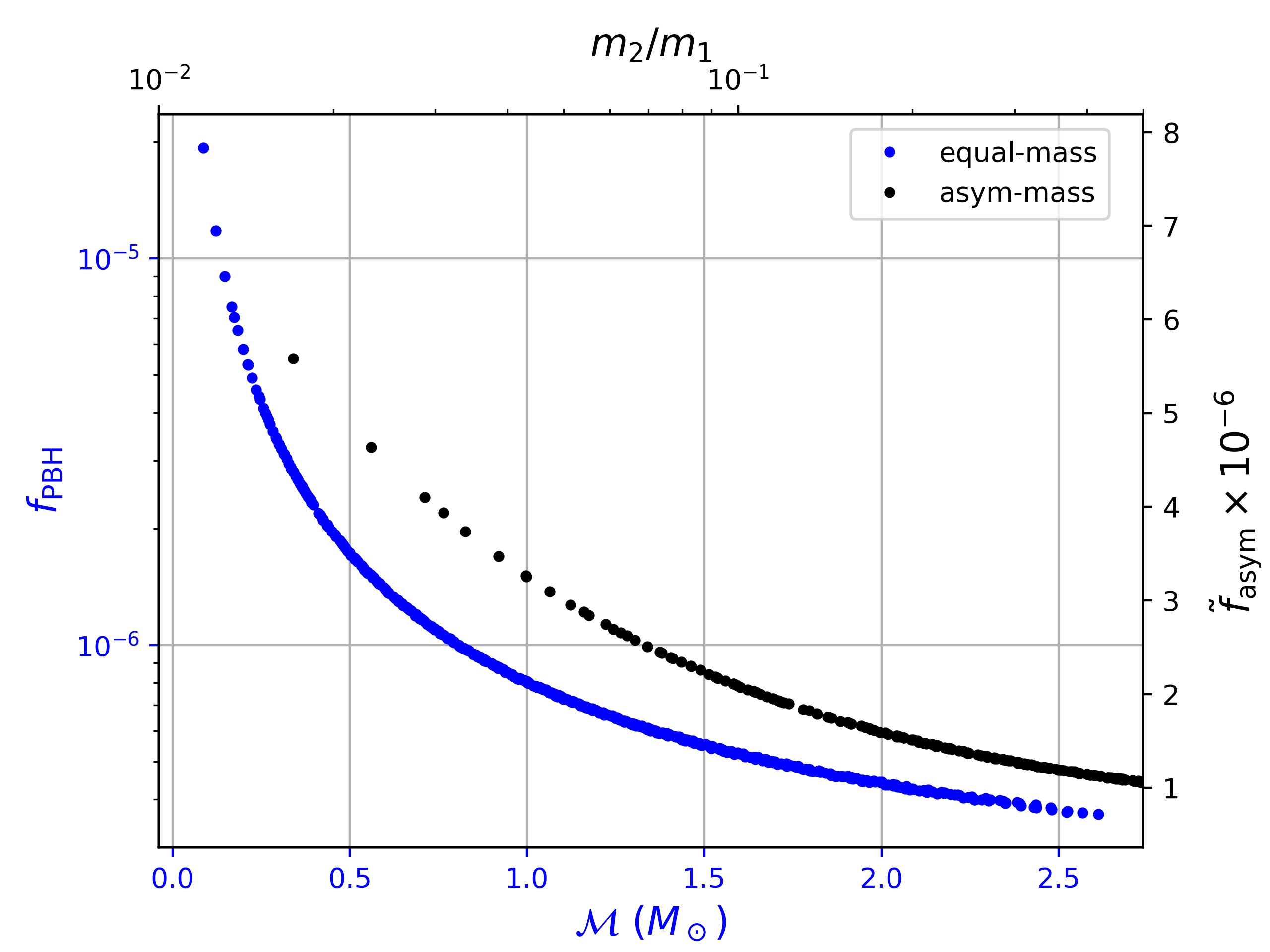}
    \caption{Projected constraints on sub-solar and solar-mass PBH binaries, for both equal-mass and asymmetric mass ratio binaries, in terms of $\fpbh$ (blue) and $\ftilde$ (black). Blue and black axes correspond to  the blue and black curve, respectively. $\fsup=2\times 10^{-3}$, and $m_1=2.5M_\odot.$}
    \label{fig:pbhconstr}
\end{figure}

\section{Conclusions} \label{sec:concl}

We have shown that the \GFH, and in general \cwh methods, can be adapted to search for long-lived compact binary inspirals in next-generation \gwh detectors. The sensitivity, considering only quasi-Newtonian orbits, has been evaluated and compared to matched filtering, and was shown to be about a factor of $\sim 4$ worse for binary neutron star masses of $\mathcal{O}(M_\odot)$. We have also quantified the maximum time available to warn astronomers of an incoming merger of two compact objects as a function of chirp mass and distance from us, ranging from a few hours to 10 minutes, and have shown that binary neutron stars with chirp masses greater than a solar mass could be detected at least 100 Mpc away from us. The sky localization, assuming that the signal's frequency evolution can be completely demodulated, was also computed, and for \bnsh systems, could be as good as 1 square degree on the sky using a single interferometer. Additionally, we have provided preliminary results detailing the robustness of our algorithm against gaps, non-stationary noise, and overlapping signals, and have quantified the optimal way to perform a semi-coherent search by carefully picking $\TFFT$ and $\Tobs$.

Our results are promising, and motivate further study of the \GFH and \cwh methods to tackle searches for compact binaries in next-generation \gwh detectors. In the case of overlapping signals, we need to quantify the impact of multiple signals on the auto-regressive power spectral density estimation, as done in \cite{Pierini:2022wgc}, revisit the choice of thresholds when constructing the time/frequency peakmap via simulations and statistical properties of the foreground/background, and determine how exactly to select candidates in the Hough plane in the presence of so many astrophysical signals. Furthermore, we must study the propagation of errors on parameters in the follow-up stages of the analysis, which will affect our ability to provide accurate sky localization to astronomers. We could also experiment with observing only the inspiral between, say, $2-3$ Hz, which would permit longer $\TFFT$, potentially allowing for more precise sky localization at the cost of some sensitivity, and finding ways to combine our sky position estimates with triangulation if the signal is seen in multiple detectors.

We must also consider the fact that the frequency evolution of the orbit may be affected by higher-order post-Newtonian terms \cite{Cutler:1994ys,Blanchet:1995ez,Poisson:1995ef,Baird:2012cu}, for which extensive waveform development has already been undertaken \cite{Pratten:2020ceb,Ossokine:2020kjp,Thompson:2020nei,Matas:2020wab}. In order to obtain parameter estimations of each of these, i.e. the symmetric mass ratio, we can envision performing hierarchical \GFH transforms. First, we obtain estimates for $f_0$ and $M_c$, demodulate the signal, and then perform successive \GFH transforms on the remaining power-law terms in the post-Newtonian expansions. It is unclear what the computational cost of this will be, and whether this is necessary, since we could hand over our estimations of source parameters to a more sensitive matched filtering pipeline at this stage. Regardless, at least an implementation of the \GFH in low-latency will be necessary to perform real-time sky localization and parameter estimation of \gwh signals from inspiraling compact binaries, and a proper comparison with matched filtering analyses in the presence of noise disturbances and overlapping signals will also be required.

\section*{Acknowledgments}
This material is based upon work supported by NSF's LIGO Laboratory which is a major facility fully funded by the National Science Foundation.

We would like to thank the Rome Virgo group for the tools necessary to perform these studies, such as the development of the original Frequency-Hough transform and the development of the short FFT databases. Additionally we would like to thank Luca Rei for managing data transfers.

We would also like to thank Thomas Dent, David Keitel, Ornella Piccinni, and Marek Szczepanczyk for their comments on the manuscript and on the science behind it. We thank Soumen Roy for discussions regarding how to run \texttt{pycbc\_geom\_nonspinbank} at one post-Newtonian order.

We thank the anonoymous referee for providing their advice to calculate the computational speed-up at at a fixed sensitivity, among their other comments, that improved the paper. 

NS is supported by grant CNS2022-135440 funded by MCIN/AEI/10.13039/501100011033 and by the European Union NextGeneration EU/PRTR. NS also acknowledges the support by the "Agence Nationale de la Recherche", grant n. ANR-19-CE31-0005-01 (PI: F. Calore).  

This research has made use of data, software and/or web tools obtained from the Gravitational Wave Open Science Center (https://www.gw-openscience.org/ ), a service of LIGO Laboratory, the LIGO Scientific Collaboration and the Virgo Collaboration. LIGO Laboratory and Advanced LIGO are funded by the United States National Science Foundation (NSF) as well as the Science and Technology Facilities Council (STFC) of the United Kingdom, the Max-Planck-Society (MPS), and the State of Niedersachsen/Germany for support of the construction of Advanced LIGO and construction and operation of the GEO600 detector. Additional support for Advanced LIGO was provided by the Australian Research Council. Virgo is funded, through the European Gravitational Observatory (EGO), by the French Centre National de Recherche Scientifique (CNRS), the Italian Istituto Nazionale della Fisica Nucleare (INFN) and the Dutch Nikhef, with contributions by institutions from Belgium, Germany, Greece, Hungary, Ireland, Japan, Monaco, Poland, Portugal, Spain.

We also wish to acknowledge the support of the INFN-CNAF computing center for its help with the storage and transfer of the data used in this paper.

We would like to thank all of the essential workers who put their health at risk during the COVID-19 pandemic, without whom we would not have been able to complete this work.

\appendix

\section{Computing $\avgVT_i$}\label{app:vt}

We provide here a quick summary of the equations we use to compute the average space-time volume in Eq. \ref{eqn:average-space-time-volume}. More details can be found in \cite{LIGOScientific:2016kwr,Taylor:2012db}.

The co-moving volume is:

\begin{equation}
    \frac{dV_c}{dz}=4\pi D_H\frac{D_L(z)^2}{(1+z)^2E(z)}.
\end{equation}
where $D_L$ is the luminosity distance, $D_H = c / H_0$, $H_0=70.4$ kms$^{-1}$Mpc$^{-1}$ is Hubble's constant, $z$ is the redshift, and:

\begin{equation}
E(z)=\sqrt{{\Omega}_{m,0}(1+z)^3+{\Omega}_{k,0}(1+z)^2+{\Omega}_{\Lambda}(z)}.
\end{equation}
% \begin{equation}
% D_H = c / H_0.
% \end{equation}
Here, the parameters for the $\Lambda$ cold dark-matter ($\Lambda$CDM) model are: $\Omega_{m,0}=0.2726$, $\Omega_{k,0}=-0.0006$, where
\cite{WMAP:2010sfg}:

\begin{equation}
\Omega_{\Lambda}(z) = \Omega_{\Lambda,0}\times {(1+z)}^{3(1+w_0+w_a)}\times e^{-3w_a\left(\frac{z}{1+z}\right)}.
\end{equation}
Here, $w_0=-1$ and $w_a=0$ are dark-energy (linear) equation of state parameters.
For different global geometries of the Universe, $D_L$, is given by:
\begin{equation}
D_L(z|\mathcal{C}) = (1+z)\times\mathcal{F}(z|\mathcal{C}),\nonumber
\end{equation}
\begin{equation}\label{eqn:lum-distance-geometry}
\mathcal{F}(z|\mathcal{C})=
\begin{cases}
 \frac{D_H}{\sqrt{\Omega_{k,0}}}\sinh\left(\sqrt{\Omega_{k,0}}\frac{D_c(z|\mathcal{C})}{D_H}\right), & \Omega_{k,0}>0, \\
 D_c(z|\mathcal{C}), & \Omega_{k,0}=0, \\
 \frac{D_H}{\sqrt{|\Omega_{k,0}|}}\sin\left(\sqrt{|\Omega_{k,0}|}\frac{D_c(z|\mathcal{C})}{D_H}\right), & \Omega_{k,0}<0,
\end{cases}
\end{equation}

The co-moving radial distance, $D_c(z)$, is given by
\begin{equation}
D_c(z) = D_H\int_0^z\frac{dz'}{E(z')},
\end{equation}

\section{Constraining binary evolution models}\label{app:pop_synth}

\citet{Belczynski:2017gds} generated a vast set of models of population I and II field binaries. The authors present a range of models and their rate densities. The rate density in each case is the result of several assumptions such as the cosmic star formation, metallicity evolution, the initial binary parameters and the implied delay time (between the birth of a binary and the final merger of two
compact objects) distribution. The details of these models are summarised in Table 2 in \citet{Belczynski:2017gds}. The data for all these models is available on the StarTrack site\footnote{http://www.syntheticuniverse.org/}. We calculate the detector frame merger rate density with our algorithm which gives the upper limit on the merger rate density in detector frame for a range of chirp masses up to redshift of $z \sim 0.08$, which we denote as $\mathcal{R}_{\rm pred}$. We compare the detector frame merger rate densities for binary neutron stars as predicted by the nine models which were not excluded by the LIGO, Virgo and KAGRA merger rate predictions (see Figs. 24 and 25 in \citet{Belczynski:2017gds}). We list the details in Table \ref{tab:evol}. We do not exclude any of these models given the constraints we estimate of the chirp masses, considering an observation time of 1 year. It would be possible to begin to exclude some of these models if we could observe for $\sim 15$ years.

\begin{table*}\label{tab:evol}
\begin{tabular}{|c|c|c|c|c|c|}
\hline
{model} &  min $\mathcal{M}$ $(M_\odot)$ &  max $\mathcal{M}$ $(M_\odot)$ &  $\mathcal{R}_{\rm pred}$ (Gpc$^{-3}$ yr $^{-1}$) & $\mathcal{R}_{95\%}$ (Gpc$^{-3}$ yr $^{-1}$) & Time to exclude (yr) \\
\hline
m13A & 0.95 &               1.72 &                                           19.89 &                                           469.66 & 23.61 \\
m23A &               0.93 &               1.70 &                                           21.07 &                                           493.95 & 23.44 \\

m30B &               0.96 &               1.66 &                                           17.08 &                                           432.18 & 25.30 \\
m33A &               0.96 &               1.70 &                                           33.19 &                                           443.36 & 13.36 \\
m40B &               0.96 &               1.67 &                                           17.01 &                                           438.23 & 25.76 \\
m43A &               0.96 &               1.70 &                                           33.84 &                                           447.99 & 13.24 \\
m50B &               0.96 &               1.66 &                                           13.09 &                                           441.49 & 33.73 \\
m60B &               0.96 &               1.66 &                                           16.52 &                                           430.75 & 26.07 \\
m70B &               0.96 &               1.67 &                                           16.69 &                                           433.60 & 25.98 \\
\hline
% \bottomrule
\end{tabular}
\caption{The binary neutron star merger rate densities predicted for different evolutionary models \cite{Belczynski:2017gds}. }
\end{table*}

\section{Computational cost calculation}
\label{app:compcost}

In order to compute the computational cost of matched filtering searches, we calculate the number of templates needed in a matched filtering analysis by running the \texttt{PyCBC} \cite{alex_nitz_2023_10137381} function \texttt{pycbc\_geom\_nonspinbank} at one post-Newtonian order for searches between $f_{\rm min,MF}=[2,11]$ Hz up to $f_{\rm max,MF}=12$ Hz for systems with $\mathcal{M}=[1,3]M_\odot$ and a minimal match of 0.97 \cite{Owen:1995tm}. The number of templates at $f_{\rm min,MF}=2,3,4,...11$ Hz was determined to be $N_{\rm temp}=$[1056637, 204110, 56928, 19988, 11974, 6649, 3389, 1544, 567, 119], and is shown in Fig. \ref{fig:temp-num}.

Assuming an observation time of one year, and calculating empirically the amount of time that each fast Fourier Transform of a template with a given number of samples takes ($\sim 0.07$ s for waveforms between 5 and 12 Hz on Apple M1 CPUs), assuming a sampling frequency of 40 Hz (for early-warning studies, this sampling frequency should be sufficient \cite{Bosi:2010glx,Sachdev:2019vvd}), we arrive at the total computational cost of a low-frequency matched-filtering search as a function of $f_{\rm min,MF}$. We then compare this computational cost to that of running the \GFH over the same parameter space by calculating the number of points in the \GFH parameter space, and timing how long each iteration of the method takes -- see \cite{Miller:2020kmv}, Sec. VC, for details on the computing cost of the \GFH.

\begin{figure}
    \centering
    \includegraphics[width=0.49\textwidth]{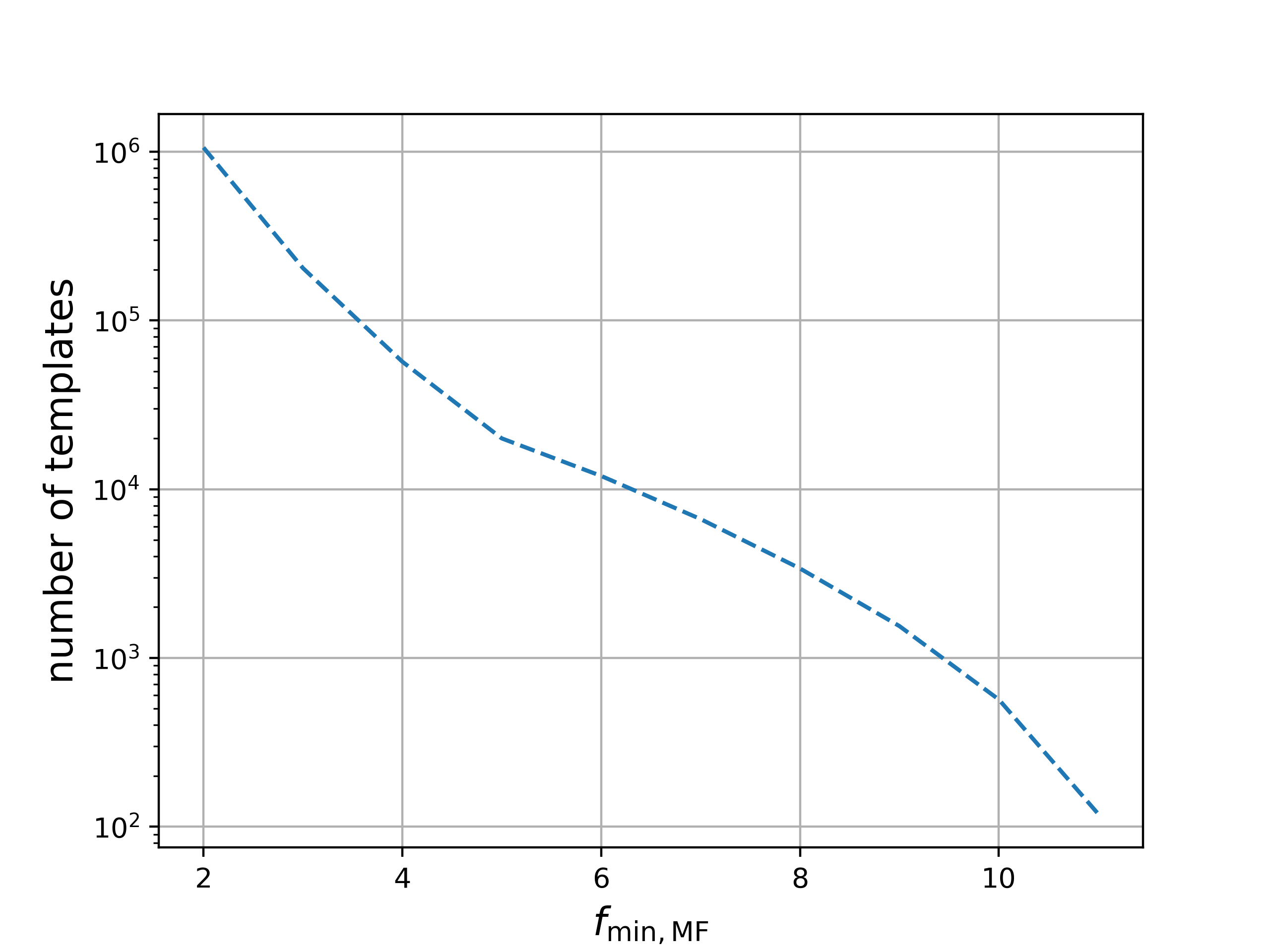}
    \caption{The number of templates in a matched filtering search for $\mathcal{M}=[1,3]M_\odot$, with a minimal match of 0.97, a minimum frequency of $f_{\rm min,MF}$ and a maximum frequency $f_{\rm max}=12$ Hz at one Post-Newtonian order.}
    \label{fig:temp-num}
\end{figure}

\bibliographystyle{apsrev4-1}
\bibliography{biblio}

\end{document}